\documentclass[a4paper,12pt]{article}
\usepackage{amsmath}
\usepackage{amssymb}
\usepackage{amsfonts}
\usepackage{mathrsfs}
\usepackage{cite}
\usepackage[font=small,labelfont=bf,position=top]{caption}
\usepackage{graphicx,subfig}
\usepackage{epstopdf}
\usepackage{lscape}

\newcommand{\vev}[1]{\left\langle{#1}\right\rangle}

\renewcommand\Re{\operatorname{Re}}
\renewcommand\Im{\operatorname{Im}}

\def\beq{\begin{equation}}
\def\eeq{\end{equation}}
\def\bea{\begin{eqnarray}}
\def\eea{\end{eqnarray}}

\def\bs{\boldsymbol}

\def\<{\left\langle}
\def\>{\right\rangle}

\newcommand{\degree}{\ensuremath{^\circ}}

\begin{document}

\title{\hfill ~\\[-30mm]
       \hfill\mbox{\small SHEP-11-29}\\[-3mm]
 \hfill\mbox{\small IPPP-11-60}\\[-3mm]
\hfill\mbox{\small DCPT-11-120}\\[13mm]
       \sffamily\Huge{Renormalisation group improved leptogenesis \\
  in family symmetry models\\[4mm]}}

\date{}

\date{}
\author{
Iain~K.~Cooper$^{1\,}$\footnote{E-mail: \texttt{ikc1g08@soton.ac.uk}}\:,~~
Stephen~F.~King$^{1\,}$\footnote{E-mail: \texttt{king@soton.ac.uk}}~~and~~
Christoph~Luhn$^{1,2\,}$\footnote{E-mail: \texttt{christoph.luhn@durham.ac.uk}}\\[9mm]
{\small\it
$^1$School of Physics and Astronomy,
University of Southampton,}\\
{\small\it Southampton, SO17 1BJ, U.K.}\\[2mm]
{\small\it
$^2$Institute for Particle Physics Phenomenology,
University of Durham,}\\
{\small\it Durham, DH1 3LE, U.K.}
}

\maketitle
\thispagestyle{empty}

\begin{abstract}\noindent
We study renormalisation group (RG) corrections relevant for leptogenesis in
the case of family symmetry models such as the Altarelli-Feruglio $A_4$ model
of tri-bimaximal lepton mixing or its extension to tri-maximal mixing.   
Such corrections are particularly relevant since in large classes of
family symmetry models, to leading order, the CP violating parameters of
leptogenesis would be identically zero at the family symmetry breaking scale,
due to the form dominance property. We find that RG corrections violate form
dominance and enable such models to yield viable leptogenesis at the scale of
right-handed neutrino masses. More generally, the results of this paper show that RG
corrections to leptogenesis cannot be ignored for any family symmetry model
involving sizeable neutrino and $\tau$ Yukawa couplings. 
\end{abstract}

\newpage
\setcounter{page}{1}

%%%%%%%%%%%%%%%%%%%%%%%%%%%%%%%%%%%%%%%%%%%%%

%%%%%%%%%%%%%%%%%%%%%%%%%%%%%%%%%%%%%%%%%%%%%

%%%%%%%%%%%%%%%%%%%%%%%%%%%%%%%%%%%%%%%%%%%%%

\setcounter{footnote}{0}

\section{Introduction}

One of the most important and well studied questions in particle physics is why the observable Universe 
has a tiny but non-zero ratio of baryons to photons without which there would be no stars, planets or life. The measurement of cosmic microwave background (CMB) anisotropies and the successful prediction of light element abundances from big bang nucleosynthesis (BBN), both lead to a consistent value of this ratio at the recombination time when atoms are formed \cite{Komatsu:2010fb},
\begin{equation}
 \label{eqn:eta}
\eta=\frac{n_B}{n_{\gamma}}\approx 6.2\times10^{-10},
\end{equation}
where $n_B$ and $n_{\gamma}$ are baryon and photon number densities respectively.\footnote{Corresponding to a portion of comoving volume containing 1 photon at temperatures where the right-handed neutrinos are relativistic.} Any theory which successfully produces such a baryon asymmetry must fulfil the famous Sakharov conditions \cite{Sakharov:1967dj} of C and CP violation, $B$ violation and departure from thermal equilibrium. One of the most popular of these is known as leptogenesis \cite{leptogenesis}, which takes advantage of the fact that non-perturbative, $B-L$ conserving, $B+L$ violating sphaleron processes can convert a lepton number asymmetry into a $B$ asymmetry. The lepton number asymmetry is obtained from the decays of heavy Majorana neutrinos and so leptogenesis is intimately linked to neutrino mass, mixing and CP violation.

The discovery of neutrino mass and mixing is arguably one of the most influential observations in particle physics in the last 15 years. It has inspired a large number of works aimed at explaining both the extremely small mass and, in particular, the striking mixing pattern (which is very different from the close to diagonal quark sector) \cite{S3-L,Dn-L,A4-L,A4-L-Altarelli,S4-L,A5-L,Dn-LQ,A4-LQ,King:2007A4,S4-LQsum,PSL-LQ,Z7Z3-LQ,delta27-LQ,delta27-LQ-Dterms,SO3-LQ,SU3-LQ,Cooper:A4typeIII,Reviews}. Seesaw mechanisms provide the most common explanation for small neutrino masses; a heavy particle is introduced which has a Yukawa coupling with the lepton doublet. When this particle is integrated out, the effective theory has a Majorana mass term for the left-handed neutrinos which is suppressed by the large mass of the new particle. This new particle must be a colour singlet but can be a weak fermionic singlet with zero hypercharge (type~I) \cite{Minkowski:1977type1}, a weak scalar triplet with two units of hypercharge (type~II) \cite{type2} or a weak fermionic triplet with zero hypercharge (type~III) \cite{Foot:1989type3}. The issue of neutrino mixing is also very well studied, and a popular technique is to postulate the existence of an extra family symmetry at high energies. This symmetry is then broken in a specific way, by the vacuum expectation value (VEV) of heavy scalars called flavons. The remnants of the breaking pattern show up in the observed mixing of neutrinos at low energies.

Many of these models of neutrino mixing (predominantly employing the type I seesaw) exhibit a property known as form dominance (FD) \cite{Chen:2009um}, defined by the condition that the columns of the neutrino Yukawa matrix are proportional to columns of the mixing matrix in a particular basis corresponding to diagonal charged lepton and right-handed neutrino mass matrices.  As discussed in several papers \cite{Choubey:2010vs,Antusch:2006cw,Jenkins:2008rb,diBari:lepto,Felipe:2009rr,AristizabalSierra:2009ex,King:2010bk}, models with family symmetry typically predict vanishing CP violating lepton asymmetry parameters~$\epsilon$ and hence zero leptogenesis.\footnote{For a discussion of how to achieve leptogenesis in the flavour symmetric phase, see e.g. \cite{AristizabalSierra:2011Arx}.} 
As pointed out in \cite{Choubey:2010vs}, this can be understood very simply from the FD property 
that the columns of the neutrino Yukawa matrix are mutually orthogonal since
they are proportional to the columns of the mixing matrix which is
unitary.\footnote{The vanishing of leptogenesis due to the orthogonality of
  the columns of the neutrino Yukawa matrix was first observed in the case of
  hierarchical neutrinos and constrained sequential dominance with
  tri-bimaximal mixing in \cite{Antusch:2006cw} and was subsequently generalised to the case of FD with any neutrino mass pattern and any mixing pattern in \cite{Choubey:2010vs}.}
However in family symmetry models the Yukawa matrices are predicted at the scale of family symmetry breaking,
which may be close to the grand unified theory (GUT) scale, and above the mass scale of right-handed neutrinos.
Therefore in such models the Yukawa matrix will be subject to renormalisation group (RG) running from the
family breaking scale down to the scale of right-handed neutrino masses relevant for leptogenesis. 
To illustrate the effects of RG corrections, we analyse two specific models involving
sizeable neutrino and $\tau$ Yukawa couplings and satisfying FD at leading order (LO):
the first model \cite{A4-L-Altarelli} reproduces the well studied
tri-bimaximal (TB) mixing pattern \cite{Harrison:2002tbm}; and the second model \cite{King:2011zj} 
reproduces the tri-maximal (TM) mixing pattern \cite{Haba:2006dz}
consistent with the results from T2K \cite{Abe:2011sj}. Although in both models RG running occurs over only one or two orders of magnitude in the energy scale, we shall show that this leads to sufficient violation of FD to allow successful leptogenesis in each case.  

One could ask why RG effects should be considered when higher order (HO) operators in the $A_4$ model have been shown to produce a realistic value of $\eta$ \cite{diBari:lepto}. The answer is that RG effects turn out to be of equal importance to HO operators in determining leptogenesis and so in general both effects should be considered together. Here we choose to drop the effect of HO operators for clarity: we want to study the effects of RG corrections to leptogenesis in isolation in order to illustrate the magnitude of the effect. Moreover, there are ultraviolet completions of the $A_4$ model of both TB \cite{Varzielas:2010mp} and TM mixing  \cite{King:2011zj} in which HO operators play a negligible role, and the viability of leptogenesis in such cases then relies exclusively on the effects of RG corrections considered here.

We emphasise that this paper represents the first study which takes into account RG corrections to leptogenesis in family symmetry models. The results in this paper show that RG corrections have a large impact on leptogenesis in any family symmetry models involving neutrino Yukawa couplings of order unity. Therefore, when considering leptogenesis in such models, RG corrections should not be ignored even when corrections arising from HO operators are also present.

The rest of the paper is organised as follows. Section \ref{sec:lepto} briefly outlines the process of calculating the baryon asymmetry of the universe $\eta$ arising from leptogenesis. Then in section \ref{sec:formdom}, the idea of FD is introduced and it is shown that the CP violating parameter in leptogenesis is indeed zero under the condition of FD. Section \ref{sec:AFModel} introduces the Altarelli-Feruglio $A_4$ model of TB neutrino mixing, while section \ref{sec:tri} introduces the parameters of the $A_4$ model of TM mixing. In section \ref{sec:RGE} we analytically estimate the RG running of the neutrino Yukawa matrices in leading log approximation. Numerical results for the baryon asymmetry of the universe arising from leptogenesis in both TB and TM models are presented in section \ref{sec:results} including contour plots of input parameters reproducing the physical value of $\eta$. Section \ref{sec:conclusion} concludes the paper.

\section{Leptogenesis}
\label{sec:lepto}
Leptogenesis takes advantage of the heavy right-handed neutrinos introduced in many models to account for the smallness of the left-handed neutrino mass. The addition of these right-handed neutrino fields $N_i$ introduces two new terms into the superpotential
\begin{equation}
\label{eqn:heavyseesaw}
W_{\nu}=\left(Y_{\nu}\right)_{\alpha i}\left(l_{\alpha}\cdot H_u\right)N_i+\frac{1}{2}N_i\left(M_{RR}\right)_{ij}N_{j},
\end{equation}
which then lead to an effective light neutrino mass once the heavy degrees of freedom are integrated out. In \eqref{eqn:heavyseesaw}, $N_i$ are the heavy right-handed neutrinos with Majorana mass matrix $M_{RR}$; $l_{\alpha}$ are the lepton doublets and $H_u$ the (hypercharge $+1/2$) Higgs doublet, which interact with $N_i$ through the Yukawa couplings $\left(Y_{\nu}\right)_{\alpha i}$.
These interactions also fulfil the well known Sakharov conditions \cite{Sakharov:1967dj} required to generate a baryon asymmetry: 1) C and CP violation (coming from the complex Yukawa coupling); 2) $B$ violation (the Majorana mass of $N$s violates $L$; sphalerons convert $\sim \frac{1}{3}$ of this into $B$ violation); 3) Departure from thermal equilibrium (due to out-of-equilibrium decays of the right-handed neutrinos). The procedure for calculation of this asymmetry is first to calculate the amount of CP violation in the decays of the right-handed neutrinos. This is then used as an input parameter to find the $B-L$ asymmetry through integration of the Boltzmann equations \cite{Buchmuller:2004nz}. These equations take into account the evolution of a $B-L$ asymmetry generated by $N$ decays against the background of 
$N$ inverse decays partially washing it out. Finally, this $B-L$ asymmetry is converted into a $B$ asymmetry using previously calculated results for sphaleron processes \cite{Khlebnikov:1988sr,Harvey:1990qw}.

\subsection{Unflavoured asymmetry}
\begin{figure*}
\centerline{
\mbox{\includegraphics[width=6in]{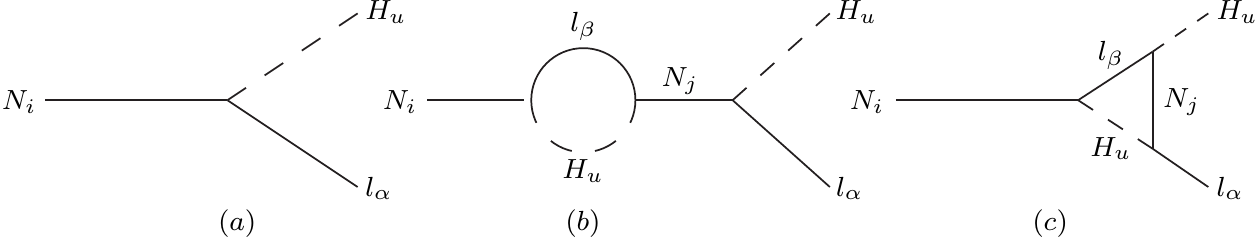}}
}
\caption{Diagrams contributing to the CP violating parameter
  $\epsilon_{i,\alpha i}$; it is the interference of $(a)$ with $(b)$ and $(c)$ which gives rise to non-zero $\epsilon_{i,\alpha i}$. Lines labelled $N$ can be any one of the seesaw particles.}
\label{fig:lepto}
\end{figure*}

To one-loop order, the CP asymmetry arises from the interference of the
diagrams in Fig.~\ref{fig:lepto}. 
Using the standard supersymmetric Feynman rules, one can calculate the decay
widths for the decay $N_i\rightarrow l_{\alpha}+H_u$,
$\Gamma_i=\sum_{\alpha}\Gamma_{\alpha i}$; these are then used to find the CP
asymmetry for $N_i$ by summing over all lepton flavours $\alpha$ \cite{Covi:Lepto},
\begin{equation}
\label{eqn:cp}
 \epsilon_i=\frac{\Gamma_i-\overline{\Gamma}_i}{\Gamma_i+\overline{\Gamma}_i}=\frac{1}{8\pi\left(Y_{\nu}^{\dagger}Y_{\nu}\right)_{ii}}\sum_{j\neq i}\mathrm{Im}\left(\left(Y_{\nu}^{\dagger}Y_{\nu}\right)_{ij}^2\right)f\left(\frac{M_j^2}{M_i^2}\right).
\end{equation}
Here, $M_i$ are the real mass eigenvalues of $M_{RR}$, and \cite{Choubey:2010vs,Antusch:2006cw,Davidson:2008bu}
\begin{equation}
 f(x_{ij})=f_{ij}=\sqrt{x_{ij}}\left(\frac{2}{1-x_{ij}}-\ln\left(\frac{1+x_{ij}}{x_{ij}}\right)\right),
\end{equation}
with $x_{ij}=\frac{M_j^2}{M_i^2}$, is the loop factor. Note that $ \epsilon_i$ is summed over all flavours of the outgoing lepton and is called the \emph{unflavoured} asymmetry. 

\subsection{Flavoured asymmetry}
\label{sub:flavoured}
The above discussion and formula for $\epsilon_i$ is relevant when the lepton doublets produced are a coherent superposition of the three flavours. This is only the case above a certain energy when the expansion rate of the universe is greater than all charged lepton interaction rates. However, as the universe cools, the $\tau$ lepton Yukawa coupling will start to come in to equilibrium at an energy of around \cite{Antusch:2006cw} $\left(1+\tan^2\beta\right)\times10^{12}$ GeV,\footnote{$\tan\beta$ here is the ratio of MSSM Higgs VEVs, and will be defined algebraically in section \ref{sec:AFModel}.} breaking the coherence of the single state superposition $e+\mu + \tau$ down into two states: the $\tau$ and the remaining coherent combination $e+\mu$. Thus, if the dynamics of leptogenesis occur below this temperature,\footnote{Strictly speaking the $\tau$ interaction rate must be faster than the $N$ inverse decay rate to overcome the Quantum Zeno effect \cite{Blanchet:2006ch}, but this is a small effect and beyond the scope of this paper.} one should take such differences into account in the calculations. The CP parameter taking into account such flavour effects is \cite{Choubey:2010vs,Antusch:2006cw,Davidson:2008bu}
\begin{equation}
\label{eqn:flavcp}
 \epsilon_{\alpha i}=\frac{1}{8\pi\left(Y^{\dagger}_{\nu}Y_{\nu}\right)_{ii}}\sum_{j\neq1}\left(\mathrm{Im}\left(Y^{*}_{\alpha i}Y_{\alpha j}(Y^{\dagger}_{\nu}Y_{\nu})_{ij}\right)f(x_{ij})+\mathrm{Im}\left(Y^{*}_{\alpha i}Y_{\alpha j}(Y^{\dagger}_{\nu}Y_{\nu})_{ji}\right)g(x_{ij})\right),
\end{equation}
with $g(x_{ij})=g_{ij}=\frac{1}{(1-x_{ij})}$ and $f_{ij}$ as above.

\subsection{Final asymmetry}
\label{sub:asymmetry}
Ultimately we want to estimate a value for the baryon to photon ratio at recombination; this is related to the $B-L$ asymmetry $N_{B-L}$ at the leptogenesis scale by \cite{DiBari:2004en}
\begin{equation}
 \eta=0.89\times10^{-2}N_{B-L}.
\end{equation}
The numerical coefficient above has two contributions: 1) from the $B-L$ conserving sphaleron processes (which are only $\sim33\%$ efficient at converting $B-L$ into $B$); 2) from scaling by photon number density in the relevant comoving volume (recall that we are calculating the baryon to photon ratio at recombination). The sphalerons convert part of the $L$ asymmetry into a $B$ asymmetry via a suppressed dimension $18$ operator active at the energies we consider, $\gg M_{EW}$. The CP asymmetries calculated in the previous section are then related to $N_{B-L}$ via
\begin{equation}
 \label{eqn:washout}
N_{B-L}=\sum_{\alpha, i}\epsilon_{\alpha i}\kappa_{\alpha i},
\end{equation}
which defines the efficiency parameters $\kappa_{\alpha i}$; these encode how efficiently the decays of $N$ produce a $B-L$ asymmetry at the leptogenesis scale. In the strong washout regime, the $\kappa_{\alpha i}$ are approximated analytically by (up to superpartner effects which increase $N_{B-L}$ by a factor of $\sqrt{2}$; see, for example, \cite{DiBari:2004en}):
\begin{equation}
 \kappa_{\alpha i}\approx\frac{2}{K_{\alpha i}z_B(K_{\alpha i})}\left(1-\exp{\left(-\frac{1}{2}K_{\alpha i}z_B(K_{\alpha i})\right)}\right),
\end{equation}
with
\begin{equation}
 z_B(K_{\alpha i})\approx 2+4(K_{\alpha i})^{0.13}\exp{\left(-\frac{2.5}{K_{\alpha i}}\right)},
\end{equation}
the decay parameter
\begin{equation}
 K_{\alpha i}=\frac{\widetilde{m}_{\alpha i}}{m^*_{MSSM}},
\end{equation}
and effective neutrino mass
\begin{equation}
 \widetilde{m}_{\alpha i}=\left(Y^{\dagger}_{\nu}\right)_{i \alpha}\left(Y_{\nu}\right)_{\alpha i}\frac{v_u^2}{M_i}.
\end{equation}
The $\widetilde{m}_{\alpha i}$ are model specific and are presented below for
the model in question (in Table~\ref{tab:Flavoured}), while
$m^*_{MSSM}=1.58\times10^{-3}\sin^2\beta$ eV \cite{Antusch:2006cw}  is the
equilibrium neutrino mass. The up-type Higgs VEV is denoted by $v_u$.
The main point to address is then the form that the Yukawa matrices take. This is discussed in the context of family symmetries \cite{S3-L,Dn-L,A4-L,A4-L-Altarelli,S4-L,A5-L,Dn-LQ,A4-LQ,King:2007A4,S4-LQsum,PSL-LQ,Z7Z3-LQ,delta27-LQ,delta27-LQ-Dterms,SO3-LQ,SU3-LQ,Cooper:A4typeIII,Reviews} which we turn to now. 

\section{Form dominance}
\label{sec:formdom}
In order to explain the observed pattern of neutrino mixing, many models
invoke the idea that a high energy family symmetry unifying the three flavours
is spontaneously broken in a specific way that leaves some imprint in the
neutrino sector at low energies. This method introduces relationships between
the parameters of $Y_{\nu}$ leading to predictions for $\epsilon_{\alpha i}$
and $\epsilon_{i}$. As discussed in the introduction, it is a striking fact
that many of these family symmetry models exhibit a property known as FD \cite{Chen:2009um}, which constrains the CP violating parameter of leptogenesis to be identically zero \cite{Choubey:2010vs}, as we now discuss.

The FD \cite{Chen:2009um} condition is that the columns of $Y_{\nu}$ 
in Eq. \eqref{eqn:heavyseesaw} are proportional to the columns of $U$,
\begin{equation}
 \label{eqn:form}
A_i=\alpha U_{i1},\quad B_i=\beta U_{i2}, \quad C_i=\gamma U_{i3},
\end{equation}
where $U$ is the unitary Pontecorvo-Maki-Nakagawa-Sakata (PMNS) matrix which is parameterised by three mixing angles and three complex phases (one Dirac and two Majorana). 
The consequences of such FD on leptogenesis is then very simple to understand: since $U$ is unitary, the columns of $Y_{\nu}$ must be mutually orthogonal. This means that the contraction $(Y^{\dagger}_{\nu}Y_{\nu})_{ij}$, 
with $i\neq j$, appearing in Eqs \eqref{eqn:cp} and \eqref{eqn:flavcp} are identically zero and so leptogenesis gives $\eta=0$. 

The FD condition also greatly simplifies the form of the effective neutrino
mass matrix arising from the  type I seesaw formula.
In terms of parameters in Eq. \eqref{eqn:heavyseesaw}, the effective neutrino mass matrix can be written,
\begin{equation}
 \label{eqn:neutrinomass}
  m_{\nu}=-v_u^2Y_{\nu}M_{RR}^{-1}Y_{\nu}^T.
\end{equation}
In the basis where the right-handed neutrinos are diagonal, i.e. that in which $M_{RR}=\mathrm{diag}\left(M_A,M_B,M_C\right)$, and writing $Y_{\nu}=\left(A,B,C\right)$, Eq. \eqref{eqn:neutrinomass} gives
\begin{equation}
 \label{eqn:columns}
m_{\nu}=-v_u^2\left(\frac{AA^T}{M_A}+\frac{BB^T}{M_B}+\frac{CC^T}{M_C}\right).
\end{equation}
In the charged lepton diagonal basis, $m_{\nu}$ is diagonalised by $U$.
Assuming FD, $m_{\nu}$ is diagonalisable independently of the parameters $\alpha,\beta,\gamma$, and, from 
(\ref{eqn:columns}) and (\ref{eqn:form}), one finds
\begin{equation}
 \label{eqn:lhnmass}
m_{\nu}^{diag}=v_u^2\mathrm{diag}\left(\frac{\alpha^2}{M_A},\frac{\beta^2}{M_B},\frac{\gamma^2}{M_C}\right).
\end{equation}
A particularly well studied case is that of TB mixing \cite{Harrison:2002tbm}. However, as emphasised in \cite{King:2010bk}, TB mixing is not linked to FD. Indeed we shall study two $A_4$ family symmetry models, one with TB mixing and one with TM mixing, where FD is present in both cases, leading to zero leptogenesis at LO, before RG corrections are included.

\section{Parameters of the ${\bs{A_4}}$ model of TB mixing}
\label{sec:AFModel}
We first consider the parameters of the Altarelli-Feruglio $A_4$ model of TB mixing
with renormalisable neutrino Dirac coupling \cite{A4-L-Altarelli},
\begin{equation}
 \label{eqn:Altarellinu}
W_{\nu}=y(lN)H_u+(x_A\xi+\widetilde{x}_A\widetilde{\xi})(NN)+x_B(\varphi_SNN),
\end{equation}
where, under $A_4$, the conventionally defined fields transform as  
$N\sim {\bf 3}$, $l\sim {\bf 3}$, $\varphi_S\sim {\bf 3}$, while $H_u\sim {\bf
  1}$ and $\xi , \tilde{\xi} \sim {\bf 1}$; the $x_i$ are constant complex parameters. 
Since this model is well known, we refer the reader to \cite{A4-L-Altarelli} for more details.
The charged lepton mass matrix in the basis used in  \cite{A4-L-Altarelli} is diagonal so the mixing structure in the neutrino sector will not receive corrections from charged lepton rotations. 
The TB structure in the neutrino sector arises from the flavon fields obtaining vacuum expectation values (VEVs) in particular directions,
\begin{equation}
 \vev{\varphi_S}=v_s\begin{pmatrix}
                  1 \\
		  1 \\
		  1
                 \end{pmatrix} , \;\vev{\xi}=u \quad \mathrm{and} \quad\vev{\widetilde\xi}=0,
\end{equation}
where the dynamics responsible for vacuum alignment has been extensively studied (for instance, in \cite{A4-L-Altarelli} for $F$-term alignment or in \cite{King:2007A4} for $D$-term alignment).

The TB structure arises in the Majorana sector of Eq. \eqref{eqn:Altarellinu}, explicitly
\begin{equation}
 \label{eqn:AFMaj}
M_{RR}=\begin{pmatrix}
            a+\frac{2b}{3} & -\frac{b}{3} & -\frac{b}{3} \\
	    -\frac{b}{3} & \frac{2b}{3} & a-\frac{b}{3} \\
	    -\frac{b}{3} & a-\frac{b}{3} & \frac{2b}{3}
           \end{pmatrix},
\end{equation}
where we define $a=2x_Au$, $b=2x_Bv_s$ as complex parameters with phase $\phi_{a,b}$. For the purposes of leptogenesis it is convenient to rotate the $N$ such that their mass matrix is diagonal.
The resulting neutrino Yukawa matrix in the diagonal $N$ basis is then, 
\begin{equation}
 \label{eqn:Altyuk}
Y_{TB}=y\begin{pmatrix}
        \frac{-2}{\sqrt{6}}e^{i\phi_A}  &  \frac{1}{\sqrt{3}}e^{i\phi_B}  &  0  \\
	\frac{1}{\sqrt{6}}e^{i\phi_A}  &  \frac{1}{\sqrt{3}}e^{i\phi_B}  &  \frac{-1}{\sqrt{2}}e^{i\phi_C}  \\
	\frac{1}{\sqrt{6}}e^{i\phi_A}  &  \frac{1}{\sqrt{3}}e^{i\phi_B}  &  \frac{1}{\sqrt{2}}e^{i\phi_C}
       \end{pmatrix}.
\end{equation}
One can see explicitly that FD is present in this model, since the columns of $Y_{TB}$ are manifestly proportional to the columns of the TB mixing matrix, 
and thus it immediately follows that $\epsilon_i=\epsilon_{\alpha i}=0$ at the scale of $A_4$ breaking. 
The phases defined in \eqref{eqn:Altyuk} are given as,
\begin{align}
\label{eqn:phases}
 \phi_A&=-\frac{1}{2}\left(\phi_{b}+\tan^{-1}\left(\frac{-|a|\sin\left(\phi_{b}-\phi_a\right)}{|b|+|a|\cos\left(\phi_{b}-\phi_a\right)}\right)\right), \\
\phi_B&=-\frac{1}{2}\phi_{a}, \\
\phi_C&=-\frac{1}{2}\left(\phi_{b}+\tan^{-1}\left(\frac{|a|\sin\left(\phi_{b}-\phi_a\right)}{|b|-|a|\cos\left(\phi_{b}-\phi_a\right)}\right)\right).
\end{align}
Therefore, there are actually only two phases ($\phi_{a}$ and $\phi_b$) and
two magnitudes ($|a|$ and $|b|$) 
%and 2 coefficients ($|x_A|$ and $|x_B|$) 
in the model, although 
%the magnitudes and coefficients always appear in pairs, e.g. $|ux_A|$, and 
only phase differences appear when considering physical quantities. This means that we may set one phase to zero without loss of generality; for our calculations we choose $\phi_a=0$.

In this basis, the Majorana neutrino mass matrix is real and diagonal and is given by
\begin{equation}
 \label{eqn:majmass}
M_{RR}^{diag}=\mathrm{diag}\left(M_1,M_2,M_3\right)=\begin{pmatrix}
         \left|a+b\right| & 0 & 0 \\
	 0 & \left|a\right| & 0 \\
	 0 & 0 & \left|-a+b\right|
        \end{pmatrix}.
\end{equation}
The effective left-handed neutrino masses are then given by\footnote{Note that in this paper we consider a normal ordering of light neutrino masses, therefore $M_1$ is the heaviest right-handed neutrino  mass. This means that $\epsilon_{3}$ and $\epsilon_{\alpha 3}$ will be dominant contributions to leptogenesis, coming from the lightest right-handed neutrino. This is simply a notational consideration, and does not affect the physics.} 
\begin{equation}
 \label{eqn:effective}
m_i=\frac{y_{\beta}^2v^2}{M_i},
\end{equation}
which incorporates the SUSY parameter $\tan\beta$ 
\begin{equation}
 \label{eqn:beta}
y_{\beta}=y\sin{\beta}\;,\quad\;\tan\beta=\frac{v_u}{v_d}\;,\quad\;v=\sqrt{v_u^2+v_d^2}=\sqrt{\left\langle H_u\right\rangle^2+\left\langle H_d\right\rangle^2}\approx174\;\mathrm{GeV}.
\end{equation}

\section{Parameters of the ${\bs{A_4}}$ model of TM mixing}
\label{sec:tri}
In light of results from T2K \cite{Abe:2011sj} indicating a sizeable reactor angle, models predicting TB mixing can potentially be ruled out. Instead, schemes such as TM mixing remain viable \cite{Haba:2006dz}:
\begin{equation}
 \label{eqn:tm}
U_{TM}=\begin{pmatrix}
        \frac{2}{\sqrt{6}}\cos\theta & \frac{1}{\sqrt{3}} & \frac{2}{\sqrt{6}}\sin\theta e^{i\rho}  \\
	-\frac{1}{\sqrt{6}}\cos\theta-\frac{1}{\sqrt{2}}\sin\theta e^{-i\rho} & \frac{1}{\sqrt{3}} & \frac{1}{\sqrt{2}}\cos\theta-\frac{1}{\sqrt{6}}\sin\theta e^{i\rho} \\
	-\frac{1}{\sqrt{6}}\cos\theta+\frac{1}{\sqrt{2}}\sin\theta e^{-i\rho} & \frac{1}{\sqrt{3}} & -\frac{1}{\sqrt{2}}\cos\theta-\frac{1}{\sqrt{6}}\sin\theta e^{i\rho}
       \end{pmatrix}.
\end{equation}
Here $\frac{2}{\sqrt{6}}\sin\theta=\sin\theta_{13}$ and $\rho$ is related to the Dirac phase. It is possible to minimally extend the Altarelli-Feruglio model above by adding a flavon in the ${\bf 1'}$ representation of $A_4$ which reproduces this pattern \cite{King:2011zj}:
\begin{equation}
 \label{eqn:oneprime}
W_{{\bf 1'}}=x_C\xi'NN ,
\end{equation}
where we define the complex parameter $c=x_C\left\langle\xi'\right\rangle$,
with phase $\phi_c$. It has been shown in \cite{King:2011zj} that the addition of this flavon doesn't affect the right-handed neutrino masses to first order, and so the parameters in common with the previous section will be unaffected. Analogously to Eq.~(\ref{eqn:Altyuk}), in the basis where charged leptons are diagonal and right-handed neutrinos are real and diagonal, the Yukawa matrix for TM mixing is,
\begin{equation}
 \label{eqn:TMYuk}
Y_{TM}\!\!=y\!\!\begin{pmatrix}
        \frac{2}{\sqrt{6}} \! & \!\! \frac{1}{\sqrt{3}} \! & \!\! \frac{2}{\sqrt{6}}\alpha_{13}^* \! \\
	-\frac{1}{\sqrt{6}}-\frac{1}{\sqrt{2}}\alpha_{13} \! & \!\!
        \frac{1}{\sqrt{3}} \! & \!\! \frac{1}{\sqrt{2}} %\cos\theta
-\frac{1}{\sqrt{6}}\alpha_{13}^* \! \\
	-\frac{1}{\sqrt{6}}+\frac{1}{\sqrt{2}}\alpha_{13} \! & \!\!
        \frac{1}{\sqrt{3}} \! & \!\! -\frac{1}{\sqrt{2}}%\cos\theta
-\frac{1}{\sqrt{6}}\alpha_{13}^* \!
	 \end{pmatrix} 
		      \!\!\begin{pmatrix}
                          \exp\left(i\phi_A\right) \!\! & \!\! 0 \!\! & \!\! 0 \! \\
			  0 \!\! & \!\! \exp\left(i\phi_B\right) \!\! & \!\! 0 \! \\
			  0 \!\! & \!\! 0 \!\! & \!\! \exp\left(i\phi_C\right) \! 
		      \end{pmatrix},
\end{equation}
where the $\phi_{A,B,C}$ are as in Eq. \eqref{eqn:phases}. We can see that the columns of this matrix are proportional to columns of $U_{TM}$ and therefore the model respects FD.
Therefore, as for the previous model of TB mixing, this model of TM mixing
also gives zero leptogenesis and $\eta=0$, to leading order.
The parameter $\alpha_{13}$ measures the deviation from TB mixing and is given
by \cite{King:2011zj} 
\begin{equation}
 \label{eqn:alpha}
\alpha_{13}=\frac{\sqrt{3}}{2}\left(\Re{\frac{c}{2\left(a-\frac{c}{2}\right)}}+\Im{\frac{c}{2\left(a-\frac{c}{2}\right)}}\frac{\Im{\frac{b}{a-\frac{c}{2}}}}{Re{\frac{b}{a-\frac{c}{2}}}}-i\frac{\Im{\frac{c}{2\left(a-\frac{c}{2}\right)}}}{\Re{\frac{b}{a-\frac{c}{2}}}}\right).
\end{equation} 

\section{Renormalisation group evolution of the Yukawa couplings}
\label{sec:RGE}

In order to generate a non-zero $\epsilon_{\alpha i}$ and $\epsilon_{i}
$, we now consider the effects of running the neutrino Yukawa couplings from the scale at which $A_4$ is broken down to the scale at which leptogenesis takes place. At one-loop, the RG equation for the neutrino Yukawa couplings in the MSSM above the scale of right-handed neutrino masses is given by \cite{King:2000hk,Chung:2003fi},
\begin{equation}
\label{eqn:RGE}
 \frac{\mathrm{d}Y_{\nu}}{\mathrm{d}t}=\frac{1}{16\pi^2}\left[N_l\cdot Y_{\nu}+Y_{\nu}\cdot N_{\nu}+\left(N_{H_u}\right)Y_{\nu}\right],
\end{equation}
where 
\begin{align}
N_l&=Y_eY_e^{\dagger}+Y_{\nu}Y_{\nu}^{\dagger}-\left(\frac{3}{2}g_2^2+\frac{3}{10}g_1^2\right)\cdot I_3, \\ N_{\nu}&=2Y_{\nu}^{\dagger}Y_{\nu}, \\ N_{H_u}&=3\mathrm{Tr}\left(Y_u^{\dagger}Y_u\right)+\mathrm{Tr}\left(Y_{\nu}^{\dagger}Y_{\nu}\right)-\left(\frac{3}{2}g_2^2+\frac{3}{10}g_1^2\right).
\end{align}
In these equations,
$t=\log\left(\frac{Q_1}{Q_0}\right)$ with $Q_1$ being the renormalisation scale and $Q_0$ the family symmetry breaking scale; $Y_{e,u}$ are the charged lepton and up-type quark Yukawa couplings respectively; $g_{1,2}$ are the\footnote{Note that $g_1$ is the GUT normalised hypercharge coupling, related to the standard hypercharge coupling $g'$ by $g_1=\sqrt{\frac{5}{3}}g'$.} ${U}(1)_Y$ and ${SU}(2)_L$ gauge couplings respectively; and $I_3$ is the $3\times3$ identity matrix. Each $N_X$ arises from all one-loop insertions allowed by gauge symmetry on the $X$-leg of the vertex.

In leading log approximation, taking the continuous derivatives to be approximately equal to a single discrete step,
Eq.~(\ref{eqn:RGE}) may be approximated as:
\begin{equation}
 \label{eqn:leadinglog}
\frac{\mathrm{d}Y_{\nu}}{\mathrm{d}t}\approx\frac{\Delta Y_{\nu}}{\Delta t}=\frac{Y_{\nu}(Q_0)-Y_{\nu}(Q_1)}{t({Q_0})-t({Q_1})}\equiv Z,
\end{equation}
yielding the solution,
\begin{equation}
\label{eqn:RGY}
Y_{\nu}(Q_1)\approx Y_{\nu}(Q_0)-Z\Delta t.
\end{equation}

As an example, we demonstrate the RG evolution of the TB Yukawa matrix in \eqref{eqn:Altyuk}
$Y_{\nu}=Y_{TB}$ (the case of $Y_{TM}$ is completely analogous). 
Inserting \eqref{eqn:Altyuk} into \eqref{eqn:RGE} and using the third family approximation then gives 
\begin{equation}
\begin{split}
 \frac{\mathrm{d}Y_{TB}}{\mathrm{d}t}&\approx\frac{y}{16\pi^2}\left(\left(J\!+\!3\left|y\right|^2\right)\begin{pmatrix}
                \frac{-2}{\sqrt{6}}e^{i\phi_A}  &   \frac{1}{\sqrt{3}}e^{i\phi_B}  &  0  \\
		\frac{1}{\sqrt{6}}e^{i\phi_A}  &  \frac{1}{\sqrt{3}}e^{i\phi_B}  &  \frac{-1}{\sqrt{2}}e^{i\phi_C}  \\
		\frac{1}{\sqrt{6}}e^{i\phi_A} &  \frac{1}{\sqrt{3}}e^{i\phi_B}  &  \frac{1}{\sqrt{2}}e^{i\phi_C}  \\
            \end{pmatrix}\right. \\
&\left. +y_{\tau}^2\begin{pmatrix}
			    0 & 0 & 0 \\
		            0 & 0 & 0 \\
                            \frac{1}{\sqrt{6}}e^{i\phi_A} &
                            \frac{1}{\sqrt{3}}e^{i\phi_B} & \frac{1}{\sqrt{2}}e^{i\phi_C}
			  \end{pmatrix}\right)\equiv Z_{TB}
\end{split}
\label{RGTB}
\end{equation}
where $J=N_{H_u}-\left(\frac{3}{2}g_2^2+\frac{3}{10}g_1^2\right)$ and $y_{\tau}$ is the Yukawa coupling of the $\tau$ lepton. This shows that the contributions from the charged lepton Yukawa couplings breaks the orthogonality of the columns, appearing as they do in only the third component of each column. This is the 
effect which gives rise to a non-zero CP violating parameter. The leading log solution for the TB case is then given by
\begin{equation}
\label{eqn:RGYTB}
Y_{TB}(Q_1)\approx Y_{TB}(Q_0)-Z_{TB}\Delta t.
\end{equation}

\section{Results}
\label{sec:results}
\begin{table}
	\centering
		\begin{tabular}{|c||c|c|}
			\hline
				 & Asymmetry & $\widetilde{m}_{\alpha i}$ \\ \hline\hline
				$\epsilon_{\alpha 1}$ & $\frac{1}{8\pi A^{\dagger}A}\left[\mathrm{Im}\left(A^*_{\alpha}B_{\alpha}\left(A^{\dagger}B\right)\right)f_{12}+\mathrm{Im}\left(A^*_{\alpha}B_{\alpha}\left(B^{\dagger}A\right)\right)g_{12}\right.$ & $\frac{\left|A_{\alpha}\right|^2}{M_1}v_u^2$
\\
 & $\quad\quad\;\,\left.+\mathrm{Im}\left(A^*_{\alpha}C_{\alpha}\left(A^{\dagger}C\right)\right)f_{13}+\mathrm{Im}\left(A^*_{\alpha}C_{\alpha}\left(C^{\dagger}A\right)\right)g_{13}\right]$ &  \\ \hline
				$\epsilon_{\alpha 2}$ & $\frac{1}{8\pi B^{\dagger}B}\left[\mathrm{Im}\left(B^*_{\alpha}A_{\alpha}\left(B^{\dagger}A\right)\right)f_{21}+\mathrm{Im}\left(B^*_{\alpha}A_{\alpha}\left(A^{\dagger}B\right)\right)g_{21}\right.$& $\frac{\left|B_{\alpha}\right|^2}{M_2}v_u^2$
\\
 & $\quad\quad\;\,\left.+\mathrm{Im}\left(B^*_{\alpha}C_{\alpha}\left(B^{\dagger}C\right)\right)f_{23}+\mathrm{Im}\left(B^*_{\alpha}C_{\alpha}\left(C^{\dagger}B\right)\right)g_{23}\right]$ & \\ \hline
				$\epsilon_{\alpha 3}$ & $\frac{1}{8\pi C^{\dagger}C}\left[\mathrm{Im}\left(C^*_{\alpha}A_{\alpha}\left(C^{\dagger}A\right)\right)f_{31}+\mathrm{Im}\left(C^*_{\alpha}A_{\alpha}\left(A^{\dagger}C\right)\right)g_{31}\right.$  & $\frac{\left|C_{\alpha}\right|^2}{M_3}v_u^2$
\\ 
& $\quad\quad\;\,\left.+\mathrm{Im}\left(C^*_{\alpha}B_{\alpha}\left(C^{\dagger}B\right)\right)f_{32}+\mathrm{Im}\left(C^*_{\alpha}B_{\alpha}\left(B^{\dagger}C\right)\right)g_{32}\right]$ & \\ \hline
		\end{tabular}
	\caption{Flavoured asymmetries and washout parameters}
	\label{tab:Flavoured}
\end{table}
In this section we present the results of our analysis for both the TB and TM models in leading log approximation.
The use of leading log approximation is justified by the small interval of energies over which the running takes place.
As before, one can represent the Yukawa matrix derived in \eqref{eqn:RGY} as $Y_{\nu}(Q_1)=\left(A(Q_1),B(Q_1),C(Q_1)\right)$ where $A(Q_1)$, $B(Q_1)$ and $C(Q_1)$ are the RG evolved versions of the column vectors in section \ref{sec:formdom}, which, as clearly seen in  \eqref{RGTB}, \eqref{eqn:RGYTB} are no longer orthogonal after RG corrections are included. 
This allows us to write the flavoured asymmetries as in Table \ref{tab:Flavoured}.
Using Eq. \eqref{eqn:flavcp} one notices 
immediately that $\epsilon_{13}=0$ since $C_1(Q_1)=0$. 
We can see that the $\epsilon_{\alpha i}$ receive a correction from RG running since, e.g. %(here the subscript $0$ represents vectors before RG evolution)
\begin{equation}
 A(Q_1)^{\dagger}B(Q_1)=\left(A(Q_0)-\left(Z\Delta t\right)_{\alpha1}\right)^{\dagger}\left(B(Q_0)-\left(Z\Delta t\right)_{\alpha2}\right),
\end{equation}
where the leading term on the right-hand side vanishes since FD implies that $A(Q_0)$ and $B(Q_0)$ (and $C(Q_0)$) are orthogonal.

In order to progress further, one will need to insert specific values for the parameters in the matrix, which are model dependent. Here, 
we make use of work presented in \cite{diBari:lepto} to fix the parameters consistently with experimental data. We take the leptogenesis 
scale $Q_1$ to be approximately the seesaw scale, $Q_1\sim (1.74^2y^2)10^{14}$ GeV (using the basic seesaw formula).\footnote{This mass scale may look quite large especially when compared to the upper bound on the reheating temperature due to the 
over-production of late-decaying gravitinos \cite{Khlopov:1984pf}. However, heavy gravitinos with masses 
$m_{3/2}>40$ TeV, will decay before nucleosynthesis. Assuming dark matter to have a significant axion/axino component, then allows reheat temperatures to be sufficiently high to produce right-handed neutrinos of mass 
$\sim 10^{14}\,\mathrm{GeV}$, as recently discussed in
e.g. \cite{Baer:2010kw}, \cite{Baer:2010gr} (and references therein).} This indicates that for small $y$ we are in the two flavour regime for $\tan \beta >10$; for larger values of $y$, $\tan\beta$ needs to be larger for us to be in the 2 flavour regime. However in the forthcoming plots, parts of the contour existing at large $y$ correspond also to larger $y_{\tau}$ and so sufficiently large $\tan\beta$. The family 
symmetry scale is around an order of magnitude below the GUT scale, roughly $Q_0 \sim 1.5\times10^{15}$ GeV; and  $y_t\sim1$. We then calculate the 
asymmetry for $0<y<2\sqrt{\pi}$ (to keep the coupling perturbative) and $0<y_{\tau}<0.5$ (to remain within bounds for $\tan\beta$ \cite{Benjamin:2010xb}).

%; this is done in both normal and inverted ordering (NO and IO).

%\begin{figure}[t]
%\label{fig:unflavoured}
%\begin{center}$
%\begin{array}{cc}
%\includegraphics[width=3in]{NOepsilons.eps} &
%\includegraphics[width=3in]{NOepsilonsvytau.eps}
%\end{array}$
%\end{center}
%\caption{Unflavoured asymmetry vs neutrino Yukawa coupling $y$ and tau lepton Yukawa coupling $y_{\tau}$}
%\end{figure}

%In the NO regime, we obtain values 
%\begin{align*}
%  &\epsilon_1= 8.59465\times10^{-8}  \\
%  &\epsilon_2= 2.28499\times10^{-7}  \\
%  &\epsilon_3=-7.81183\times10^{-6}
%\end{align*}
%using values of $y=3$ and $y_{\tau}=0.5$; the fact that $y_{\tau}$ is so large means that we should be taking into account of the %flavour of the outgoing lepton since $10^{9}\left(1+\tan^2\beta\right)GeV\lesssim %M_1\lesssim10^{12}\left(1+\tan^2\beta\right)GeV$. This regime is known as the two-flavour regime since the $\tau$-lepton Yukawa %coupling is in thermal equilibrium while the $e$- and $\mu$- Yukawas are indistinguishable, meaning we have two effective %'flavours' to consider, $e+\mu$ and $\tau$.

%The IO regime is slightly more complicated since the masses $M_1$ and $M_2$ are very close to being degenerate; we thus have to consider whether $(M_1-M_2)^2\ll M_1^2\Gamma_2^{(0)2}$ or vice versa in order to decide whether to use the resonant propagator or not. In fact it turns out that for $y\gtrsim0.877$, we should take account of the decay width of the internal neutrino;

\begin{figure}[t]
\begin{center}$
\begin{array}{cc}
\includegraphics[width=3.3in]{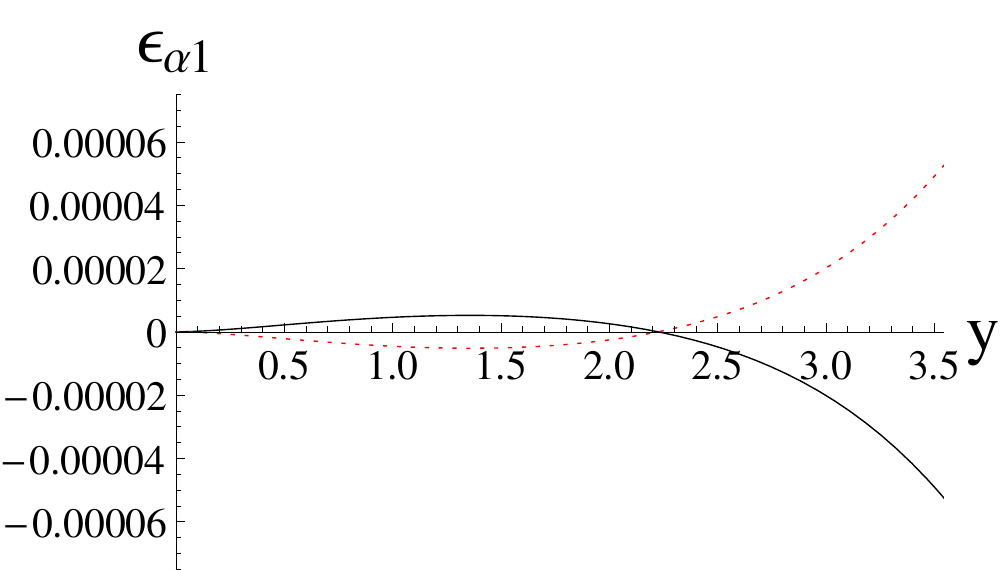} &
\includegraphics[width=3.3in]{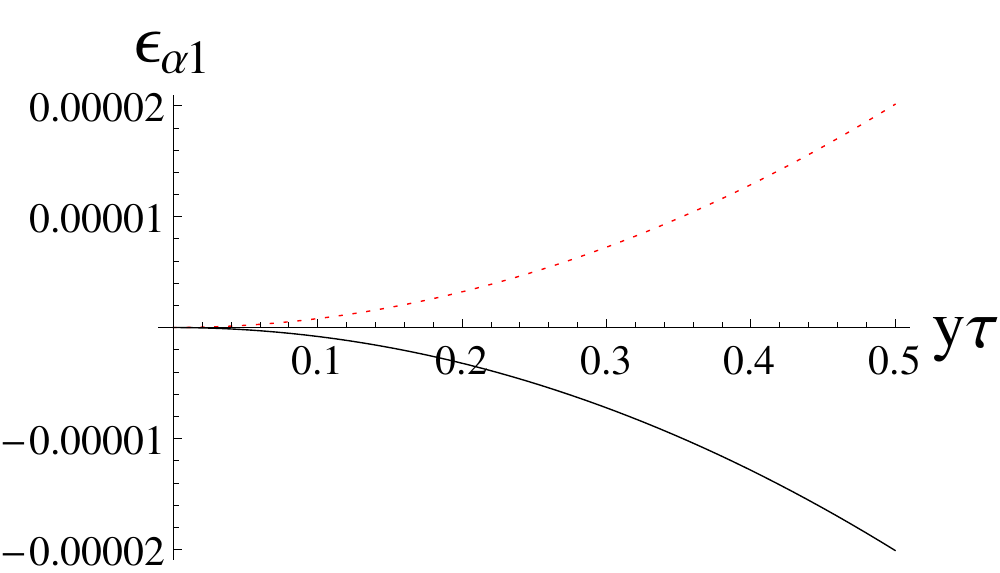} \\
\includegraphics[width=3.3in]{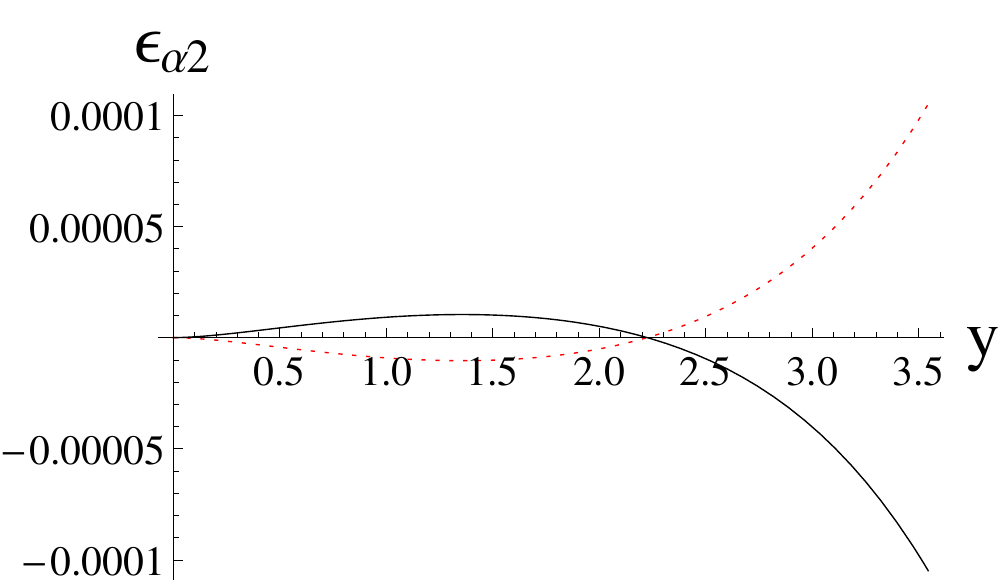} &
\includegraphics[width=3.3in]{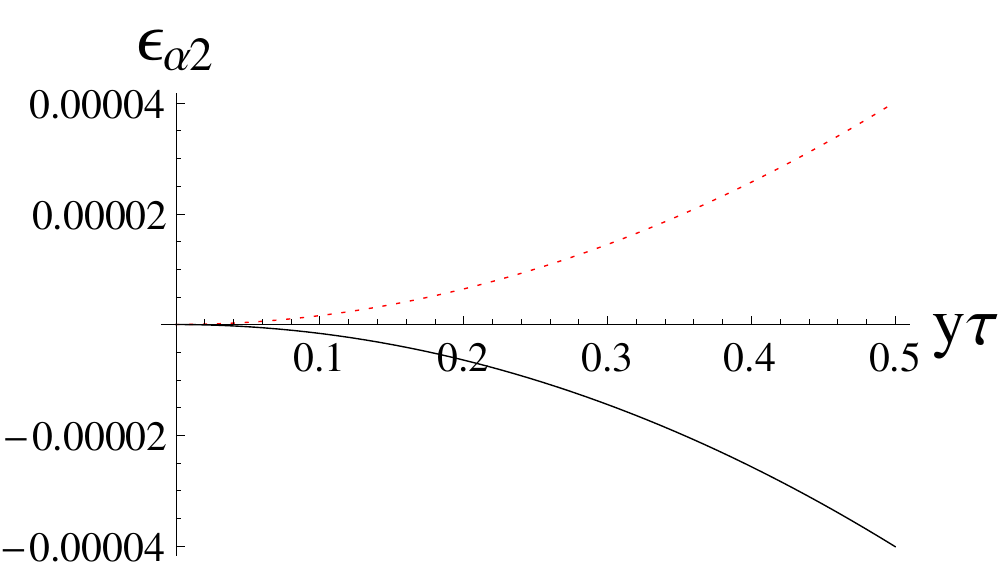} \\
\includegraphics[width=3.3in]{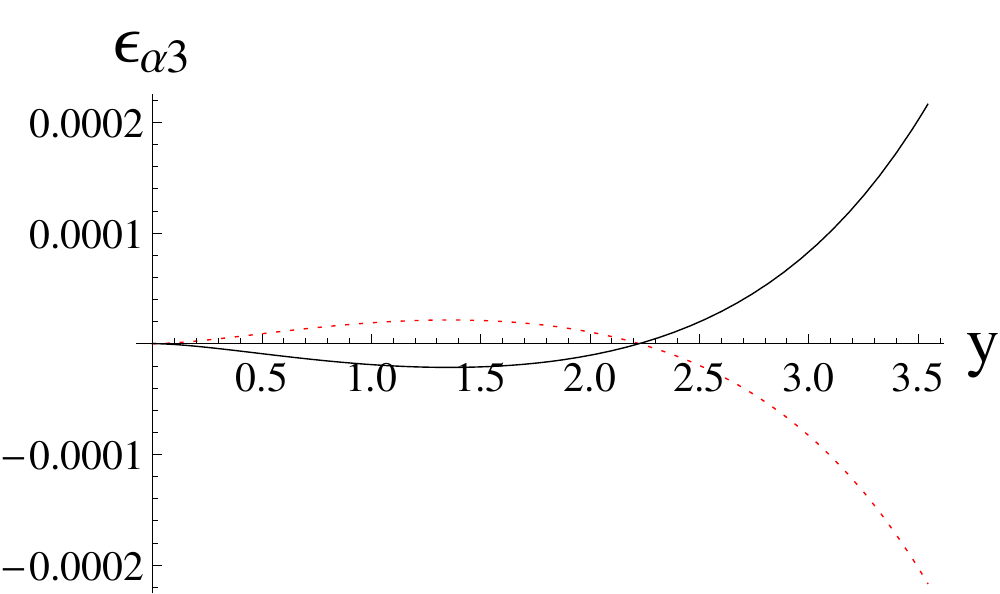} &
\includegraphics[width=3.3in]{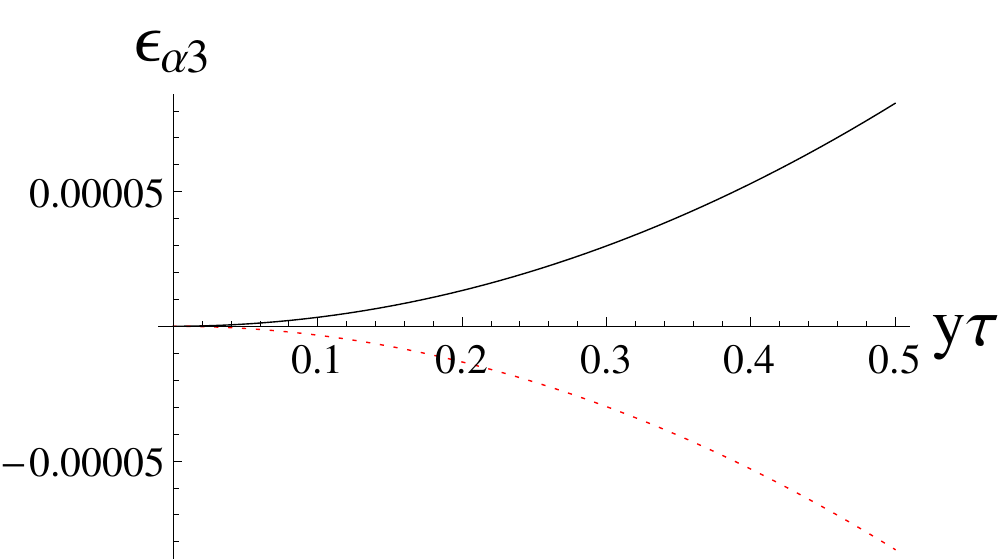}
\end{array}$
\end{center}
\caption{Flavoured asymmetries plotted against neutrino Yukawa $y$ and tau lepton Yukawa $y_{\tau}$ in the two flavour regime (i.e. $e-\mu$ and $\tau$) for the TB model. In the $y_{\tau}$ graphs, $y$ is fixed to be $3$, while in the $y$ graphs, $y_{\tau}$ is fixed to be $0.5$. $\epsilon_{e\mu, i}$ are black solid lines while $\epsilon_{\tau,i}$ are red dashed lines.}
\label{fig:flavouredgraphs}
\end{figure}

\subsection{TB mixing}
\label{sub:TB}
We now specialise to the case of RG improved leptogenesis in the TB model, where the
TB Yukawa couplings are given in (\ref{eqn:RGYTB}), repeated below,
\begin{equation}
\label{eqn:RGYTB2}
Y_{TB}(Q_1)\approx Y_{TB}(Q_0)-Z_{TB}\Delta t.
\end{equation}
The results for the flavoured asymmetries versus $y$ and $y_{\tau}$ are presented in Fig.~\ref{fig:flavouredgraphs}, in the two-flavour regime. It can be seen that the contributions from $\epsilon_{\alpha 3}$ are the dominant ones, as expected.

\begin{figure}[t]
\begin{center}
\includegraphics[width=6in]{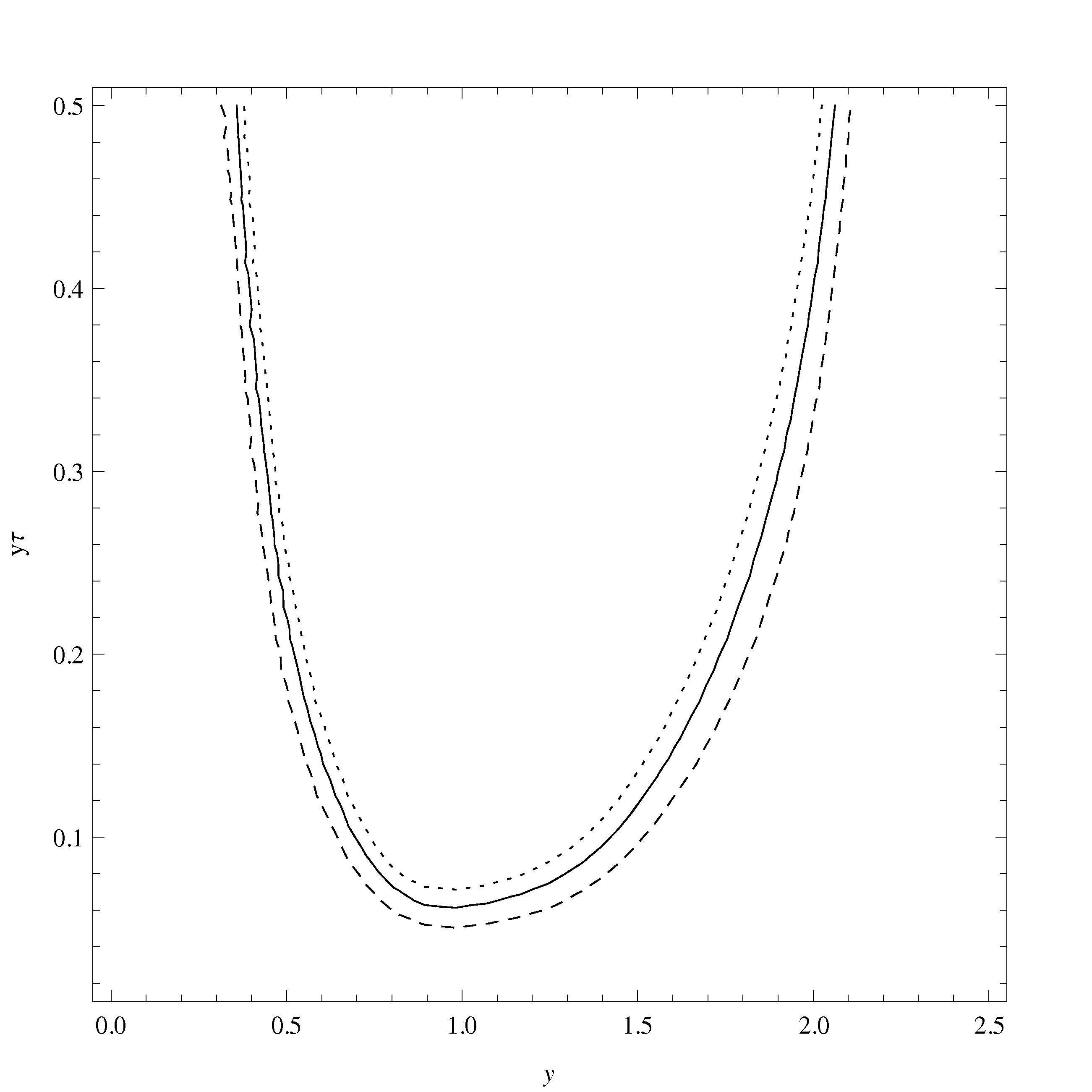}
\end{center}
\caption{A plot showing the contours of the baryon to photon ratio $\eta$ in the tau Yukawa, $y_{\tau}$, versus neutrino Yukawa, $y$, plane. The dotted and dashed lines are $\eta=8.2\times10^{-10}$ and $\eta=4.2\times10^{-10}$ while the solid line is the measured value of $\eta=6.2\times10^{-10}.$}
\label{fig:expcontour}
\end{figure}

Following the procedure set out in section \ref{sub:asymmetry}, we then calculate the baryon to photon ratio $\eta$. Fig. \ref{fig:expcontour} displays the contour matching the experimentally measured value of $6.2\times10^{-10}$, along with two others, demonstrating the sensitivity of the required Yukawa couplings to the value of $\eta$. This shows that there is a definite range of Yukawa couplings for which a realistic matter-antimatter asymmetry can be obtained purely by considering RG evolution of the neutrino Yukawa matrix, without the need for any extra particles or HO operators to be considered.

\subsection{TM mixing}
\label{sub:TM}
\begin{figure}[ht]
\begin{center}$
\begin{array}{cc}
\includegraphics[width=3.3in]{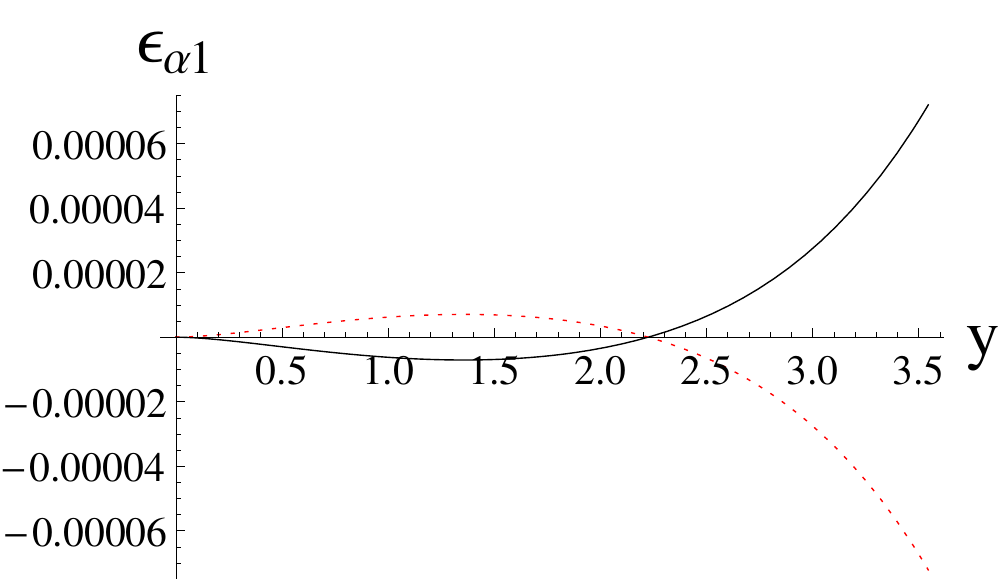} &
\includegraphics[width=3.3in]{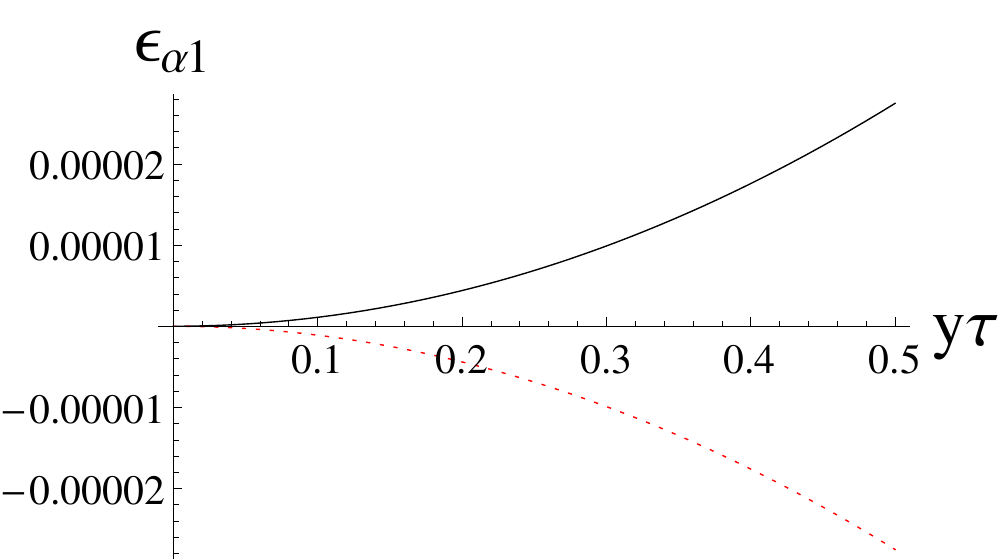} \\
\includegraphics[width=3.3in]{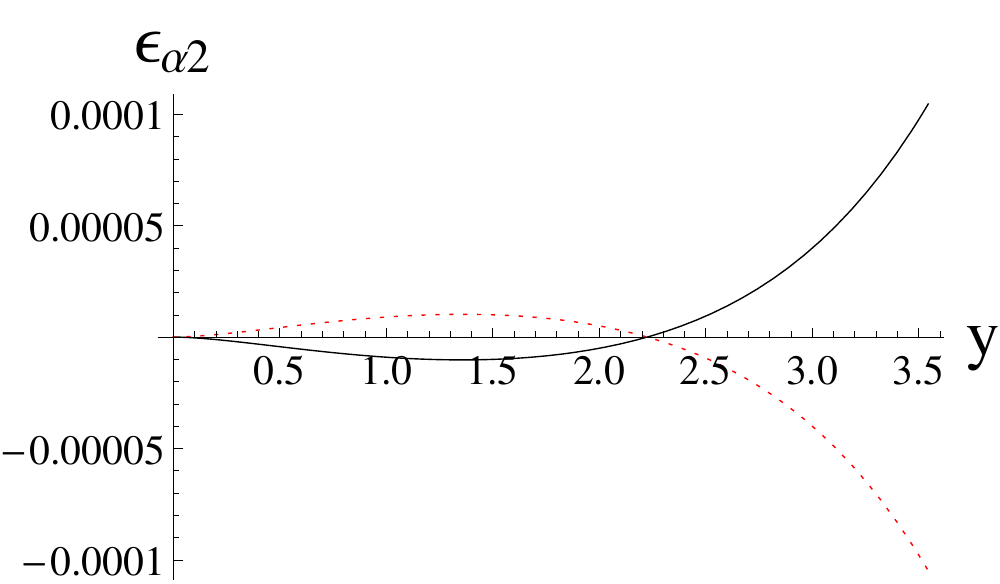} &
\includegraphics[width=3.3in]{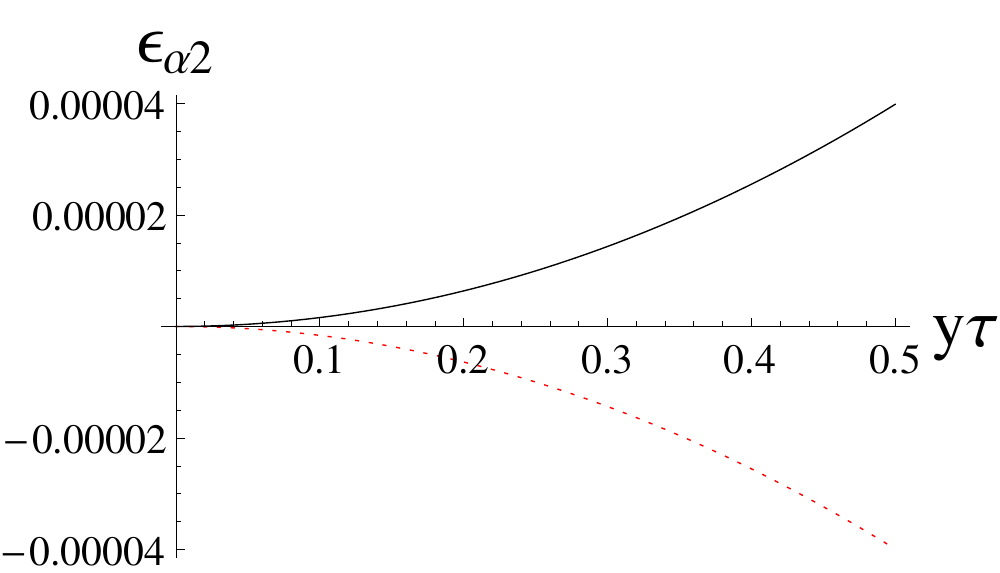} \\
\includegraphics[width=3.3in]{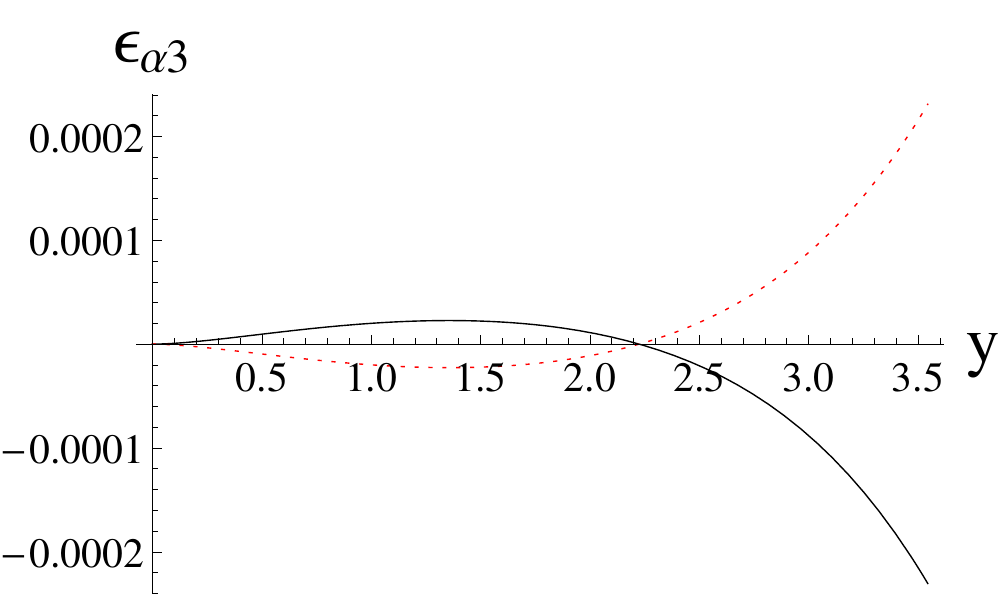} &
\includegraphics[width=3.3in]{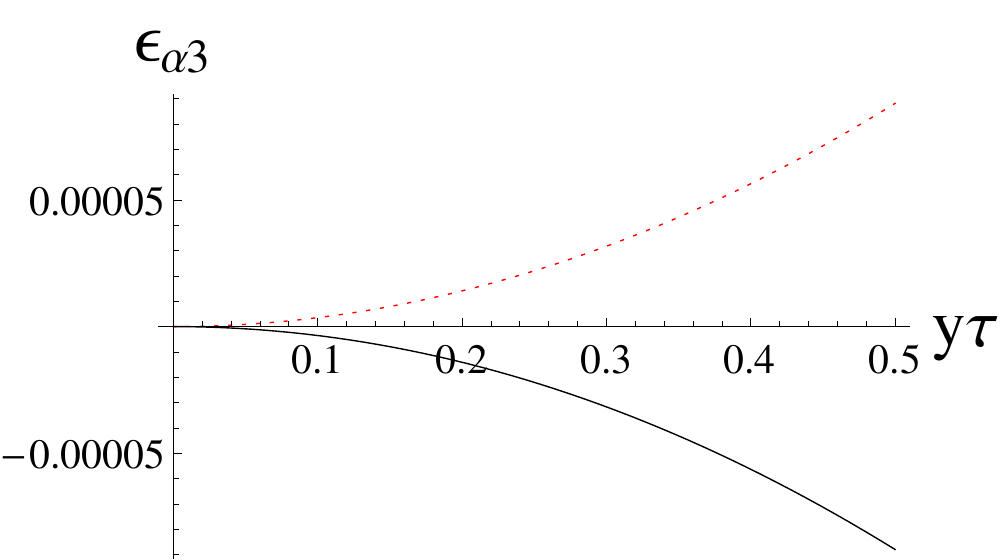}
\end{array}$
\end{center}
\caption{Flavoured asymmetries plotted against neutrino Yukawa $y$ and tau
  lepton Yukawa $y_{\tau}$ in the two flavour regime (i.e. $e-\mu$ and $\tau$)
  for the TM model with $\theta_{13}=8\degree$ and a real parameter $c=x_C
  \langle \xi'\rangle$. In the $y_{\tau}$ graphs, $y$ is fixed to be $3$, while in the $y$ graphs, $y_{\tau}$ is fixed to be $0.5$. $\epsilon_{e\mu,i}$ are black solid lines while $\epsilon_{\tau,i}$ are red dashed lines.}
\label{fig:TMflavouredgraphs}
\end{figure}

\begin{figure}[ht]%
$\begin{array}{cc}
\subfloat[][$\phi_c=0 \; \mathrm{rad}$]{\includegraphics[width=3.3in]{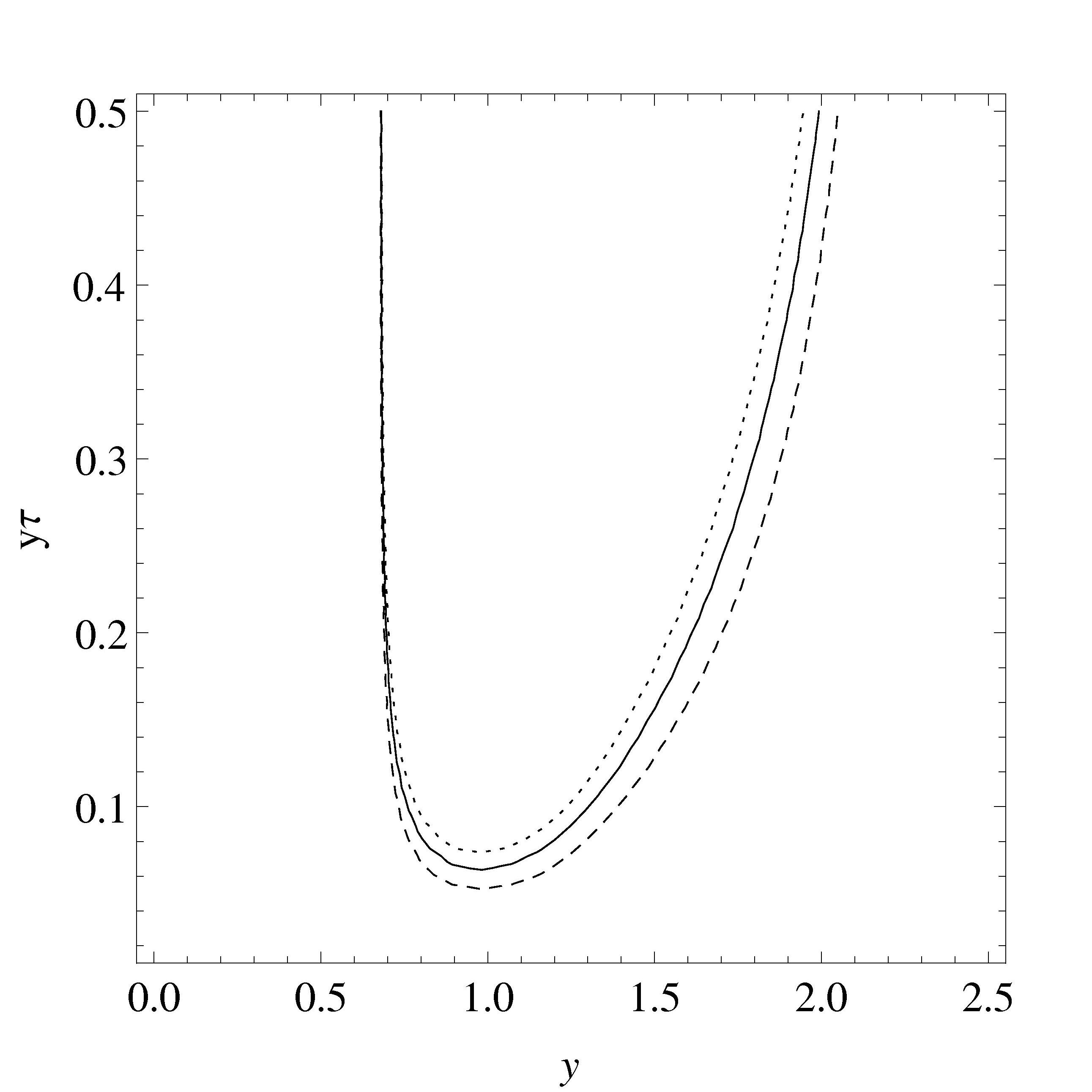}}%
&
\subfloat[][$\phi_c=0.91 \; \mathrm{rad}$]{\includegraphics[width=3.3in]{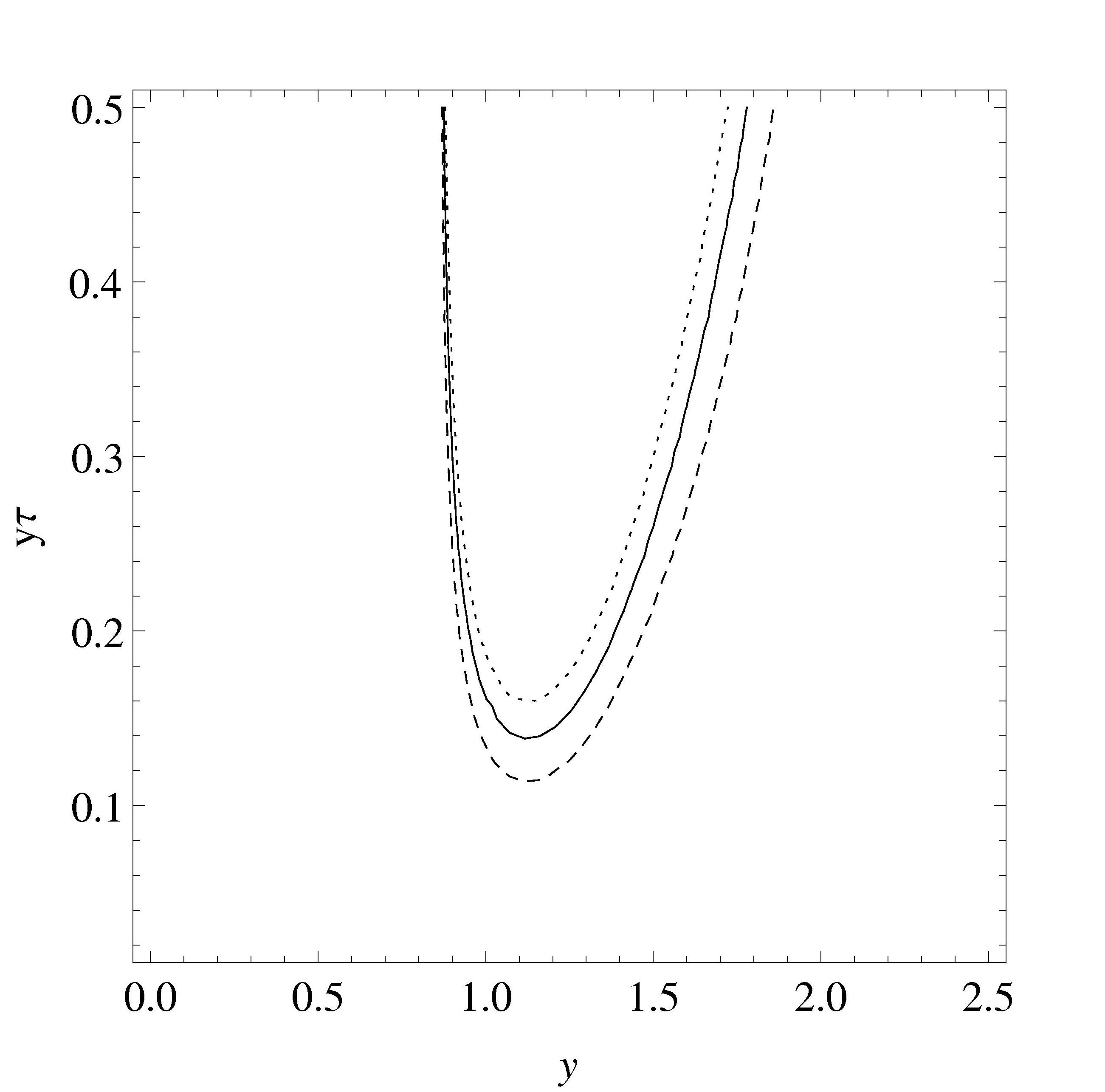}}%
\\
\subfloat[][$\phi_c=3.28 \; \mathrm{rad}$]{\includegraphics[width=3.3in]{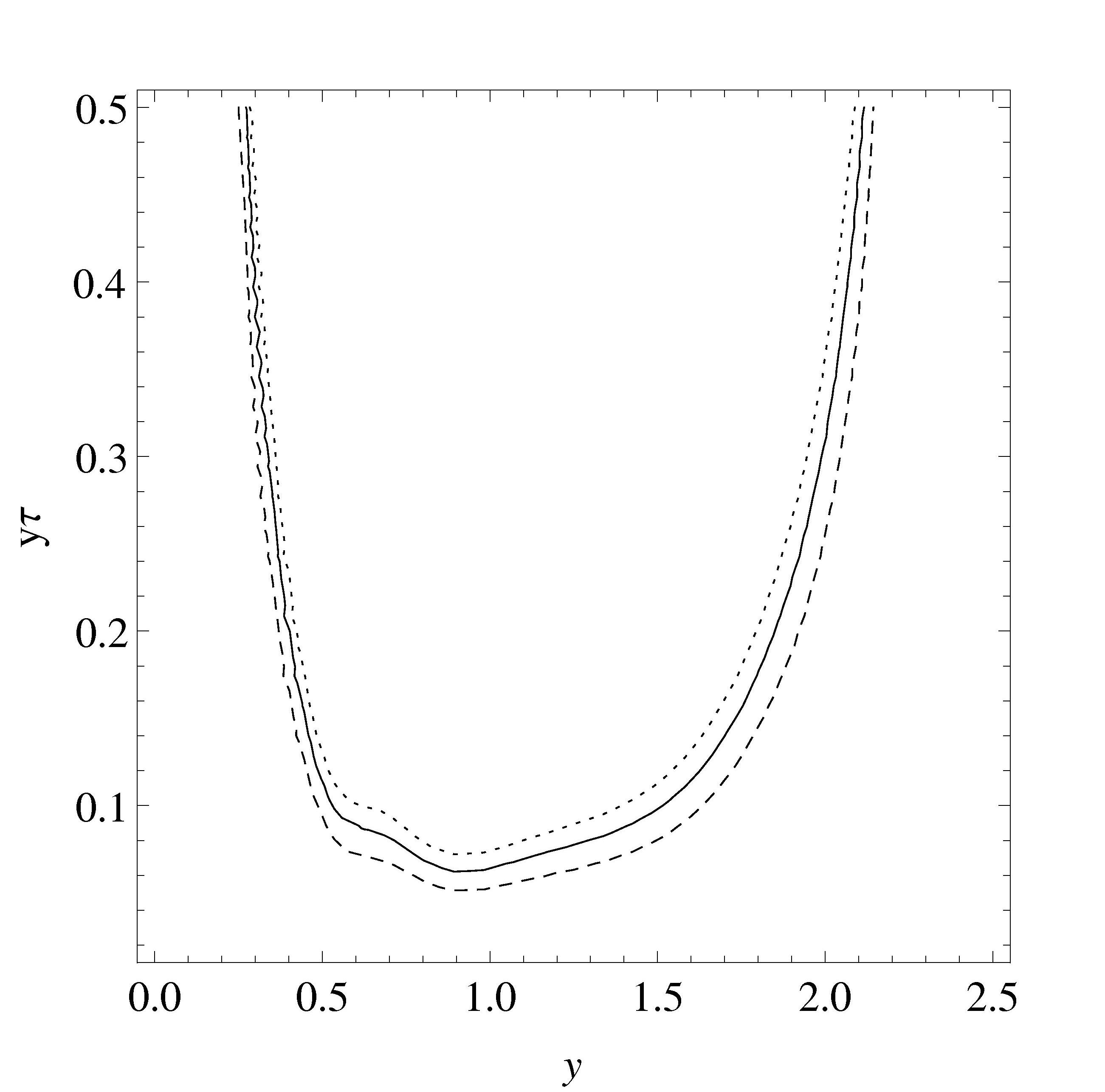}}%
&
\subfloat[][$\phi_c=6 \; \mathrm{rad}$]{\includegraphics[width=3.3in]{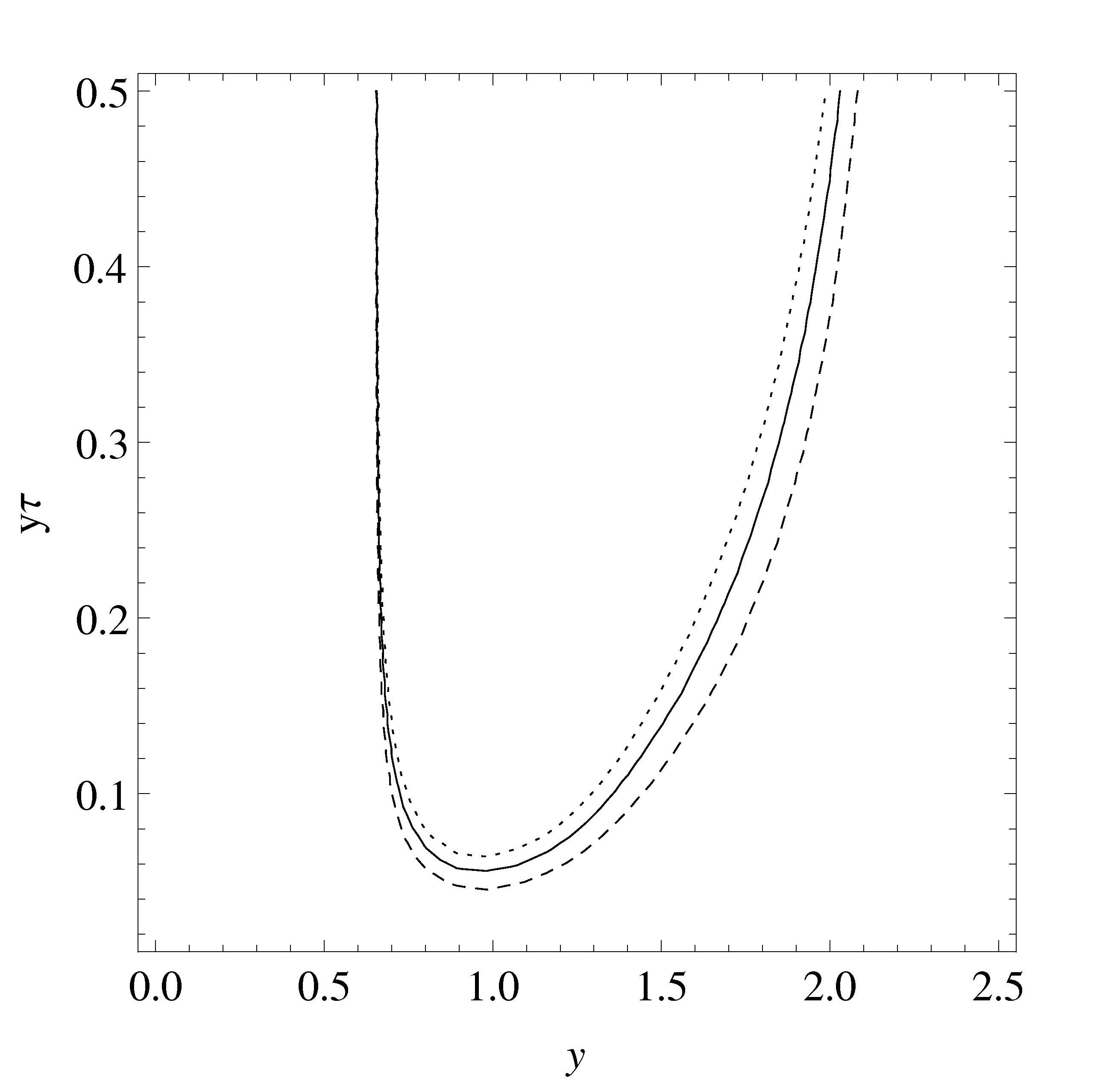}}%
\end{array}$
\caption{These plots show contours of the baryon to photon ratio $\eta$ from
  the TM model with $\theta_{13}=8\degree$ in the tau Yukawa, $y_{\tau}$,
  versus neutrino Yukawa, $y$, plane. The dotted and dashed lines are
  $\eta=8.2\times10^{-10}$ and $\eta=4.2\times10^{-10}$ while the solid line
  is the measured value of $\eta=6.2\times10^{-10}$. Each plot is for a
  different value of the phase of $c=x_C \langle \xi'\rangle$, given above the relevant panel.}%}
\label{fig:TMcontourscentral}%
\end{figure}

\begin{figure}[ht]%
$\begin{array}{cc}
\subfloat[][$\phi_c=0 \; \mathrm{rad}$]{\includegraphics[width=3.3in]{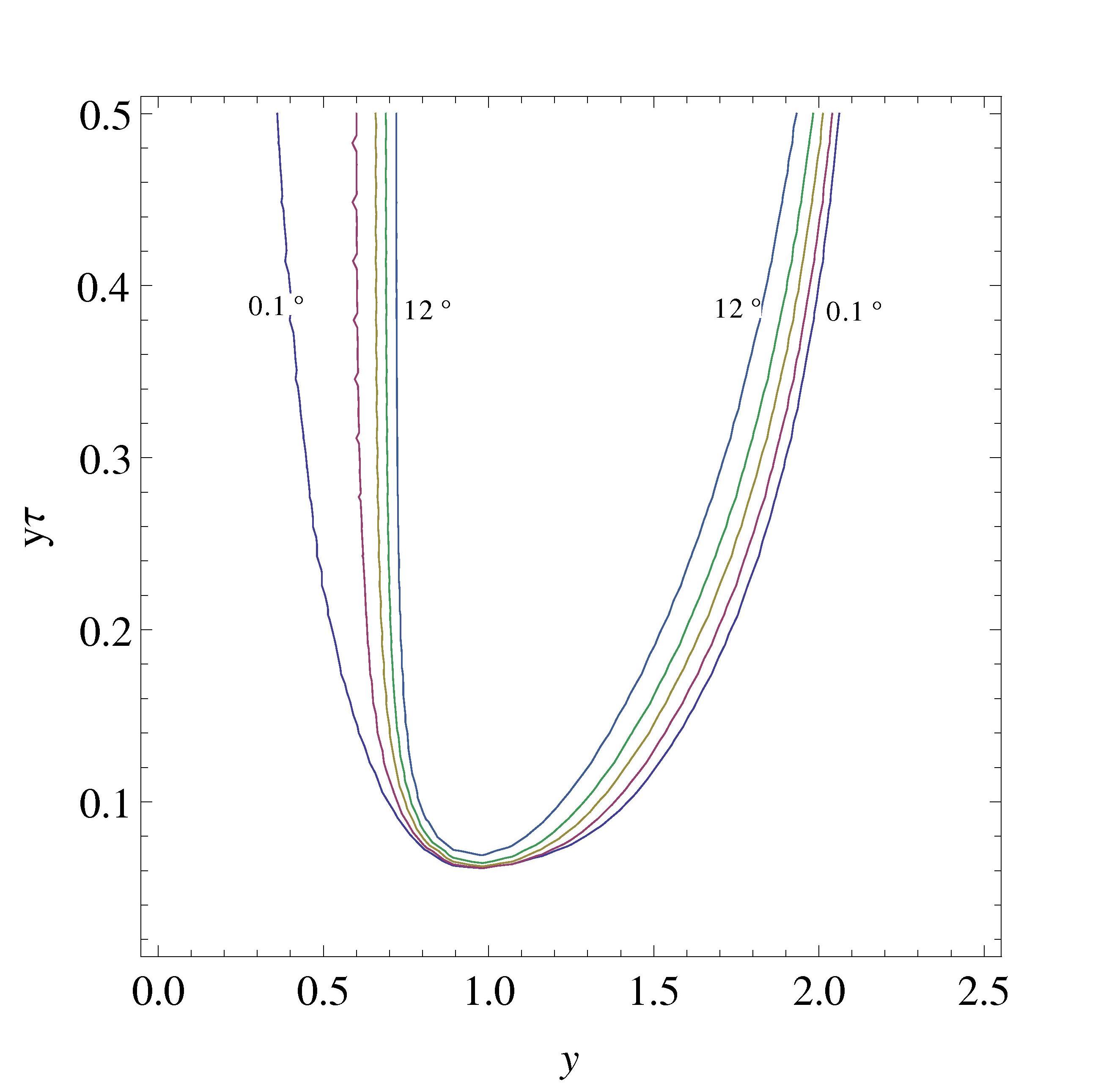}}%
&
\subfloat[][$\phi_c=0.91 \; \mathrm{rad}$]{\includegraphics[width=3.3in]{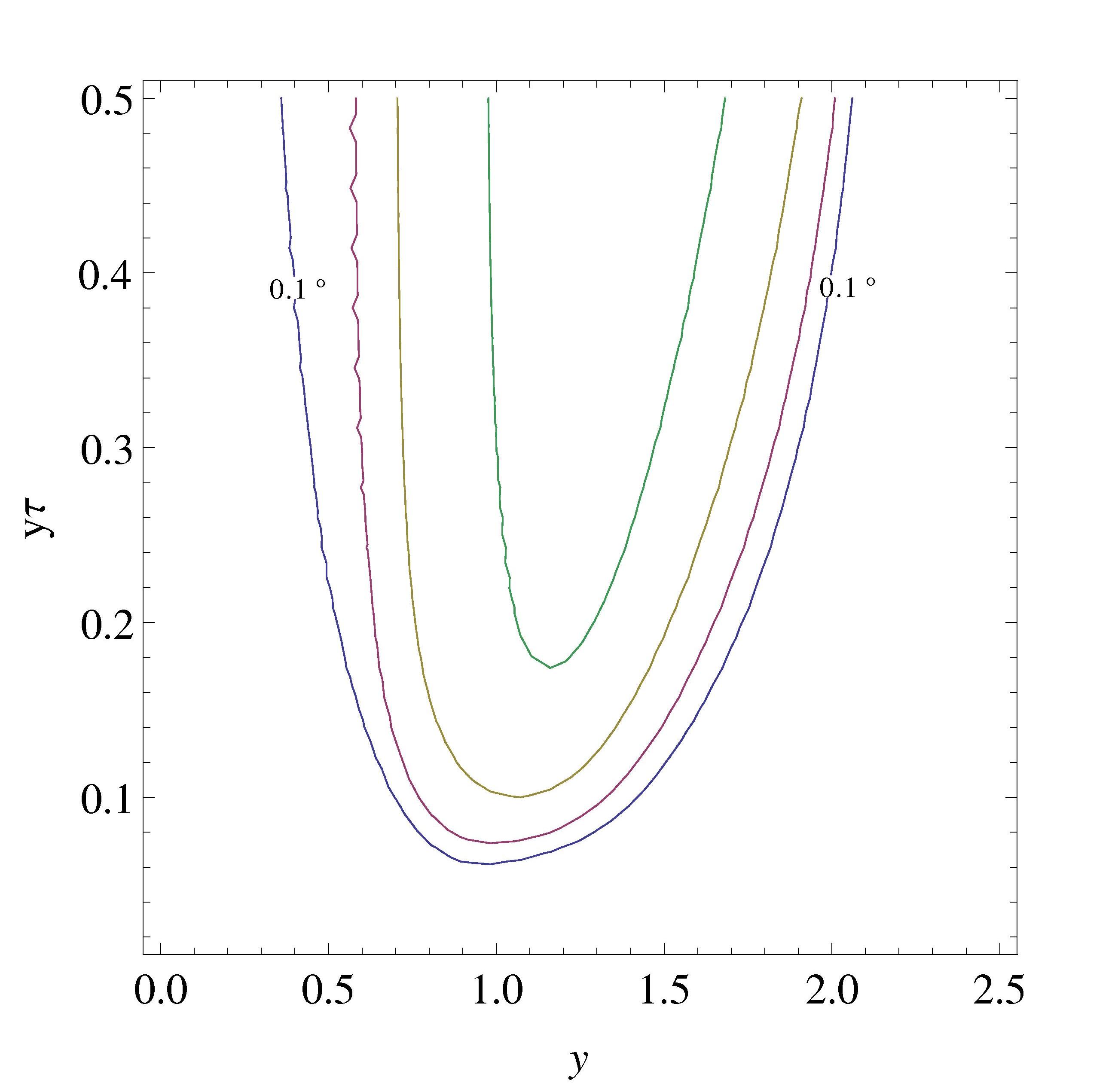}}%
\\
\subfloat[][$\phi_c=3.28 \; \mathrm{rad}$]{\includegraphics[width=3.3in]{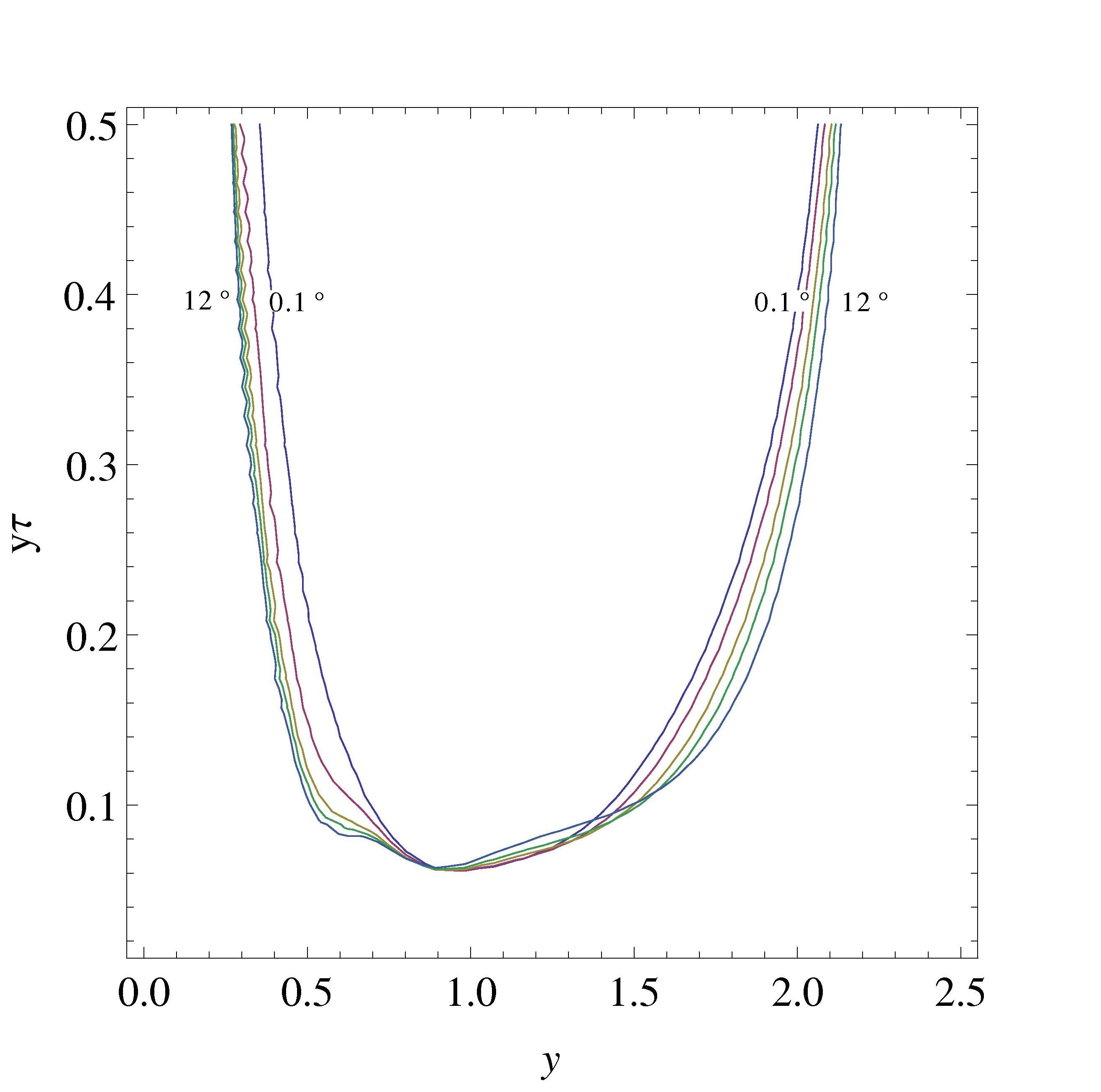}}%
&
\subfloat[][$\phi_c=6 \; \mathrm{rad}$]{\includegraphics[width=3.3in]{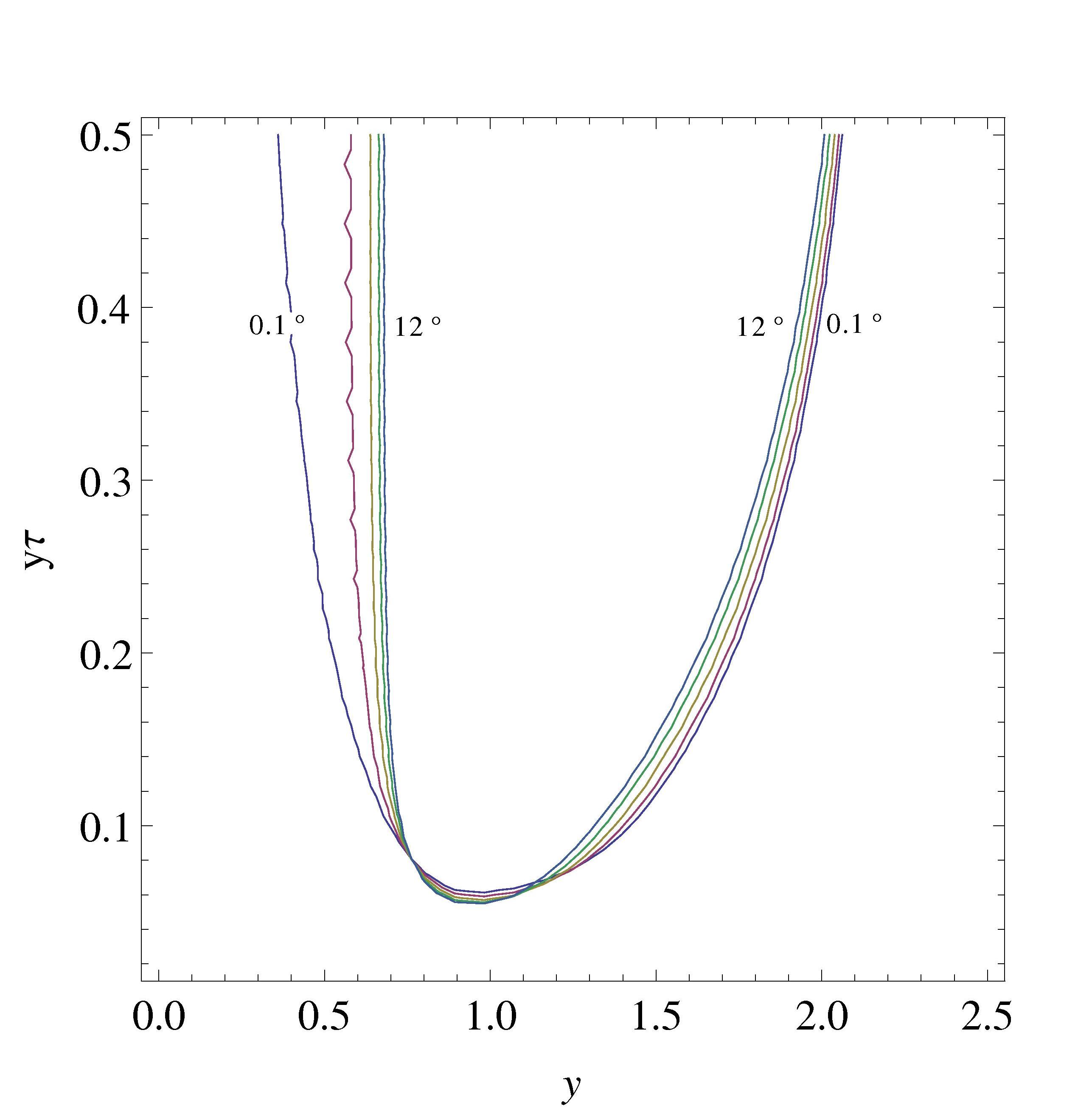}}%
\end{array}$
\caption{These plots show contours of the baryon to photon ratio
  $\eta=6.2\times10^{-10}$ from the TM model with
  $\theta_{13}=0.1\degree,3\degree,6\degree,9\degree,12\degree$ (purple, red,
  yellow, green, blue respectively) in the tau Yukawa, $y_{\tau}$, versus
  neutrino Yukawa, $y$, plane. Each plot is for a different value of the phase
  of $c=x_C \langle\xi' \rangle$, given above the relevant panel. Note that the $\theta_{13}=12\degree$ contour is not possible for 
$\phi_c=0.91$ radians.}%}
\label{fig:TMcontoursoverlay}%
\end{figure}

We now perform a similar analysis in the TM model, using the RG improved Yukawa matrix 
analogous to (\ref{eqn:RGYTB}), namely,
\begin{equation}
\label{eqn:RGYTM}
Y_{TM}(Q_1)\approx Y_{TM}(Q_0)-Z_{TM}\Delta t.
\end{equation}
where the high energy Yukawa matrix $Y_{TM}(Q_0)$ is given in (\ref{eqn:TMYuk}), 
with $Z_{TM}$ analogous to (\ref{RGTB})
and otherwise assuming similar parameters to the case of TB mixing.
However one must choose the new complex parameter $c$ carefully in order to satisfy the relation \cite{King:2011zj}
\begin{equation}
\label{eqn:theta13}
 \frac{\sqrt{6}}{2}\sin{\theta_{13}}=\left|\alpha_{13}\right|.
\end{equation}
We present flavoured asymmetries for $\theta_{13}=8\degree$ (the current T2K central value \cite{Abe:2011sj}) and $\phi_c=0$ in Fig. \ref{fig:TMflavouredgraphs}. We also plot contours of $\eta=4.2\times10^{-10},\;6.2\times10^{-10},\;8.2\times10^{-10}$ for $\theta_{13}=8\degree$ and $\eta=6.2\times10^{-10}$ for $\theta_{13}=0.1\degree,\;3\degree,\;6\degree,\;9\degree,\;12\degree$; and for each value of $\theta_{13}$, we present four different choices of phase and modulus of $c$ which satisfy \eqref{eqn:theta13}. These can be seen in Figs  \ref{fig:TMcontourscentral} and \ref{fig:TMcontoursoverlay}. For small $\theta_{13}$ and therefore small $c$, the results are very similar to those for TB mixing (c.f. Figs \ref{fig:expcontour} and \ref{fig:TMcontoursoverlay} purple line), which is expected since the only difference between the two models is the presence of the $\xi'$ flavon. For the larger values of $\theta_{13}$, it is clear that changing $c$ has a significant effect as one can see from the variation of contours in Fig. \ref{fig:TMcontoursoverlay}; for instance the $12\degree$ contour for a phase of $\phi_c=0.91$ rad doesn't show up across the whole displayed plane. 

%It is interesting to note the splitting of the main contour into two lobes in figure \ref{fig:TMcontours}; one might worry that the edges of each lobe look like discontinuous cliffs. However, figure \ref{fig:etavy} shows that $\eta$ is continuous, it merely changes sign in the region between the two lobes. 

%\begin{figure}[t]
%\begin{center}
%\includegraphics[width=6in]{contours.eps}
%\end{center}
%\caption{A plot showing the contours of the baryon to photon ratio $\eta$ in the neutrino Yukawa, $y$, versus tau Yukawa, %$y_{\tau}$, plane.}
%\label{fig:contours}
%\end{figure}

\section{Conclusion}
\label{sec:conclusion}

In this paper we have studied RG corrections relevant for leptogenesis in the case of 
family symmetry models such as the Altarelli-Feruglio
$A_4$ model of tri-bimaximal lepton mixing or its extension to tri-maximal mixing.  
Such corrections are particularly relevant since in large classes of family
symmetry models, to leading order, the CP violating parameters of leptogenesis
would be identically zero at the family symmetry breaking scale, due to the
form dominance property.  We have used the third family approximation, keeping only the largest Yukawa couplings,
subject to the constraint of perturbativity. In addition, 
the $\tau$ Yukawa coupling is related to the SUSY parameter $\tan\beta$, which has had experimental bounds placed upon it. 

Our results demonstrate that it is possible to obtain the observed value for
the baryon asymmetry of the Universe in models with FD by exploiting RG
running of the neutrino Yukawa matrix over the small energy interval between
the family symmetry breaking scale and the right-handed neutrino mass scale
$\sim 10^{14}\,\mathrm{GeV}$.  
Of course, the importance of RG corrections applies more generally than to the particular models we have considered here for illustrative purposes, and the right-handed neutrino masses may be lower in some models.

In conclusion, the results in this paper show that RG corrections have a large impact on leptogenesis in any family symmetry models involving neutrino and charged lepton Yukawa couplings of order unity, even though the range of RG running between the flavour scale and the leptogenesis scale may be only one or two orders of magnitude in energy. Therefore, when considering leptogenesis in such models, RG corrections should not be ignored, even when corrections arising from HO operators are also present.

\section*{Acknowledgements}
We thank Pasquale Di Bari and David A. Jones for discussions regarding
leptogenesis calculations throughout this work. The authors acknowledge
support from the STFC Rolling Grant No. ST/G000557/1.

%%%%%%%%%%%%%%%%%%%%%%%%%%%%%%%%%%%%%%%%%%%%%%%%%%%%%%%%%%%%%%%%%%%%%%%%%%%%%%%

%%%%%%%%%%%%%%%%%%%%%%%%%%%%%%%%%%%%%%%%%%%%%%%%%%%%%%%%%%%%%%%%%%%%%%%%%%%%%%%

%%%%%%%%%%%%%%%%%%%%%%%%%%%%%%%%%%%%%%%%%%%%%%%%%%%%%%%%%%%%%%%%%%%%%%%%%%%%%%%

\providecommand{\bysame}{\leavevmode\hbox to3em{\hrulefill}\thinspace}

\end{document}